# Towards Practical Implementation of Deep Random Secrecy

Thibault de Valroger [(*)]

Abstract

We have formerly introduced Deep Random Secrecy, a new cryptologic technique capable to ensure secrecy as close as desired from perfection against unlimited passive eavesdropping opponents. We have also formerly introduced an extended protocol, based on Deep Random Secrecy, capable to resist to unlimited active MITM. The main limitation of those protocols, in their initial presented version, is the important quantity of information that needs to be exchanged between the legitimate partners to distill secure digits. We have defined and shown existence of an absolute constant, called Cryptologic Limit, which represents the upper-bound of Secrecy rate that can be reached by Deep Random Secrecy protocols. At last, we have already presented practical algorithms to generate Deep Randomness from classical computing resources. This article is presenting two optimization techniques, the first one is based on recombination and reuse of random bits, the second is obtained by adapting sampling parameters depending on the public information published by the partners. These techniques enable to dramatically increase the bandwidth performance of formerly introduced protocols, without jeopardizing the entropy of secret information. That optimization enables to envision an implementation of Deep Random Secrecy at very reasonable cost. The article also summarizes former results in the perspective of a comprehensive implementation.

## I. Introduction and summary of former work

Modern cryptography mostly relies on mathematical problems commonly trusted as very difficult to solve, such as large integer factorization or discrete logarithm, belonging to complexity theory. No certainty exists on the actual difficulty of those problems. Some other methods, rather based on information theory, have been developed since early 90's. Those methods relies on hypothesis about the opponent (such as « memory bounded » adversary [6]) or about the communication channel (such as « independent noisy channels » [5]) ; unfortunately, if their perfect secrecy have been proven under given hypothesis, none of those hypothesis are easy to ensure in practice. At last, some other methods based on physical theories like quantum indetermination [3] have been described and experimented, but they remain complex to implement.

*(*) See contact and information about the author at last page*

Considering this theoretically unsatisfying situation, we have proposed in [9] to explore a new path, where proven information theoretic security can be reached, without assuming any limitation about the capacities of the opponent, who is supposed to have unlimited computation and storage power, nor about the communication channel, that is supposed to be perfectly public, accessible and equivalent for any playing party (legitimate partners and opponents). Furthermore, while we were only considering passive unlimited opponents in [9], we consider in this work active unlimited MITM opponents.

In our model of security, the legitimate partners of the protocol are using Deep Random generation to generate their shared encryption key, and the behavior of the opponent, when inferring secret information from public information, is governed by Deep Random assumption, that we introduce. In active opponent scenarios, the legitimate partners have an initial shared authentication secret, that is used only for authentication purpose, not for generating the shared encryption key.

**Back on the Deep random assumption**

We have introduced in [9] the Deep Random assumption, based on Prior Probability theory as developed by Jaynes [7]. Deep Random assumption is an objective principle to assign probability, compatible with the symmetry principle proposed by Jaynes [7].

Before presenting the Deep Random assumption, it is needed to introduce Prior probability theory.

If we denote $\mathfrak{I}_<$ the set of all prior information available to observer regarding the probability distribution of a certain random variable $X$ ('prior' meaning before having observed any experiment of that variable), and $\mathfrak{I}_>$ any public information available regarding an experiment of $X$, it is then possible to define the set of possible distributions that are compatible with the information $\mathfrak{I} \triangleq \mathfrak{I}_< \cup \mathfrak{I}_>$ regarding an experiment of $X$; we denote this set of possible distributions as:

$$D_{\mathfrak{I}}$$

The goal of Prior probability theory is to provide tools enabling to make rigorous inference reasoning in a context of partial knowledge of probability distributions. A key idea for that purpose is to consider groups of transformation, applicable to the sample space of a random variable $X$, that do not change the global perception of the observer. In other words, for any transformation $\tau$ of such group, the observer has no information enabling him to privilege $\varphi_{\mathfrak{I}}(v) \triangleq P(X = v|\mathfrak{I})$ rather than $\varphi_{\mathfrak{I}} \circ \tau(v) = P(X = \tau(v)|\mathfrak{I})$ as the actual conditional distribution. This idea has been developed by Jaynes [7].

We will consider only finite groups of transformation, because one manipulates only discrete and bounded objects in digital communications. We define the acceptable groups $G$ as the ones fulfilling the 2 conditions below:

$(C1)$ Stability - For any distribution $\varphi_{\mathfrak{I}} \in D_{\mathfrak{I}}$, and for any transformation $\tau \in G$, then $\varphi_{\mathfrak{I}} \circ \tau \in D_{\mathfrak{I}}$

$(C2)$ Convexity - Any distribution that is invariant by action of $G$ does belong to $D_{\mathfrak{I}}$

It can be noted that the set of distributions that are invariant by action of $G$ is exactly:

$$R_{\mathfrak{I}}(G) \triangleq \left\{ \frac{1}{|G|} \sum_{\tau \in G} \varphi_{\mathfrak{I}} \circ \tau \,|\forall \varphi_{\mathfrak{I}} \in D_{\mathfrak{I}} \right\}$$

The set of acceptable groups as defined above is denoted:

$\Gamma_{\mathfrak{I}}$

For any group $G$ of transformations applying on the sample space $F$, we denote by $\Omega_{\mathfrak{I}}(G)$ the set of all possible conditional expectations when the distribution of $X$ courses $R_{\mathfrak{I}}(G)$. In other words:

$$\Omega_{\mathfrak{I}}(G) \triangleq \{Z(\mathfrak{I}) \triangleq E[X|\mathfrak{I}] | \forall \varphi_{\mathfrak{I}} \in R_{\mathfrak{I}}(G)\}$$

Or also:

$$\Omega_{\mathfrak{I}}(G) = \left\{Z(\mathfrak{I}) = \int_F v\varphi_{\mathfrak{I}}(v)dv \,|\forall \varphi_{\mathfrak{I}} \in R_{\mathfrak{I}}(G)\right\}$$

The **Deep Random assumption** prescribes that, if $G \in \Gamma_{\mathfrak{I}}$, the strategy $Z_\xi$ of the opponent observer $\xi$, in order to estimate $X$ from the public information $\mathfrak{I}$, should be chosen by the opponent observer $\xi$ within the restricted set of strategies:

$$\boldsymbol{Z_\xi \in \Omega_{\mathfrak{I}}(G)} \qquad (\boldsymbol{A})$$

The Deep Random assumption can thus be seen as a way to restrict the possibilities of $\xi$ to choose his strategy in order estimate the private information $X$ from his knowledge of the public information $\mathfrak{I}$. It is a fully reasonable assumption because the assigned prior distribution should remain stable by action of a transformation that let the distribution uncertainty unchanged.

$(A)$ suggests of course that $Z_\xi$ should eventually be picked in $\bigcap_{G\in\Gamma_{\mathfrak{I}}} \Omega_{\mathfrak{I}}(G)$, but it is enough for our purpose to find at least one group of transformation with which one can apply efficiently the Deep Random assumption to the a protocol in order to measure an advantage distilled by the legitimate partners compared to the opponent.

**Back on the presentation of protocol $\mathcal{P}$ (introduced in [9])**

The following protocol has been presented in [9]. In order to shortly remind the notations, let's set $x = (x_1, \dots, x_n)$ and $y = (y_1, \dots, y_n)$ some parameter vectors in $[0,1]^n$ and $i = (i_1, \dots, i_n)$ and $j = (j_1, \dots, j_n)$ some Bernoulli experiment vectors in $\{0,1\}^n$, we denote :

$x.y$ (resp. $i.j$) the scalar product of $x$ and $y$ (resp. $i$ and $j$)

$$|x| \triangleq \sum_{s=1}^n x_s \; ; \; |i| \triangleq \sum_{s=1}^n i_s$$

$\forall \sigma \in \mathfrak{S}_n$, $\sigma(x)$ represents $(x_{\sigma(1)}, \dots, x_{\sigma(n)})$

$\frac{x}{k}$ represents $\left(\frac{x_1}{k}, \dots, \frac{x_n}{k}\right)$ for $k \in \mathbb{R}_+^*$

In that protocol, besides being hidden to any third party (opponent or partner), the probability distributions used by each legitimate partner also need to have specific properties in order to prevent the opponent to efficiently evaluate $V_A$ by using internal symmetry of the distribution.

Those specific properties are:

(i) Each probability distribution $\Phi$ (for $A$ or $B$) must be « far » from its symmetric projection $\bar{\Phi}(x) \triangleq \frac{1}{n!}\sum_{\sigma\in\mathfrak{S}_n} \Phi\circ\sigma(x)$

(ii) At least one of the distribution (of $A$ or $B$) must avoid to have brutal variations (Dirac)

The technical details explaining those constraints are presented in [9]. The set of compliant distributions is denoted $\zeta(\alpha)$ where $\alpha$ is a parameter that quantifies the « remoteness » of a distribution from its symmetric projection.

For such a distribution $\Phi$, a tidying permutation, denoted $\sigma_\Phi$, is a permutation that realizes the minimum (or maximum):

$$\min_{\sigma\in\mathfrak{S}_n}\left(\sum_{r,s\in I_0\times\bar{I}_0}\int_{[0,1]^n} x_{\sigma(r)}x_{\sigma(s)}\Phi(x)dx\right)$$

where $I_0 = \{1, \dots, {}^{n}/_{2}\}$. Minimizing the above criteria is an efficient way to ensure that $\Phi\circ\sigma_\Phi$ is driven away from its symmetric projection. Again, details are presented in [9].

Here are the steps of the proposed protocol:

$A$ and $B$ are the legitimate partners. The steps of the protocol $\mathcal{P}(\alpha, n, k, K, L)$ are the followings:

***Step 1 – Deep Random Generation****: A and B use their respective DRG to pick independently the respective probability distributions $\Phi$ and $\Phi' \in \zeta(\alpha)$. $\Phi$ (resp. $\Phi'$) is then secret (under Deep Random assumption) for any observer other than A (resp. B) beholding all the published information. A (resp. B) calculates a tidying permutation $\sigma_\Phi$ (resp. $\sigma_{\Phi'}$) of $\Phi$ (resp. $\Phi'$). A draws the parameter vector $x \in \{0,1\}^n$ from $\Phi$. B draws the parameter vector $y \in \{0,1\}^n$ from $\Phi'$.*

***Step 2 – Degradation****: $k > 1$ is a public Degradation parameter; A generates a Bernoulli experiment vectors $i \in \{0,1\}^n$ from the parameter vector $x/k$. A publishes i. B generates a Bernoulli experiment vectors $j \in \{0,1\}^n$ from the parameter vector $y/k$. B publishes j.*

***Step 3 – Dispersion:*** *A and B also pick a second probability distribution respectively $\Psi$ and $\Psi' \in \zeta(\alpha)$ such that it is also secret (under Deep Random assumption) for any observer other than A (resp. B). $\Psi$ is selected also such that $\int_{|x|\in[k|i|-\sqrt{n},k|i|+\sqrt{n}]}\Psi(x)dx \geq \frac{1}{2\sqrt{n}}$ in order to ensure that $|i|$ is not an unlikely value for $\approx |x/k|$ (same for $\Psi'$ by replacing x by y and i by j). $\Psi$ (resp. $\Psi'$) is used to scramble the publication of the tidying permutation of A (resp. B). A (resp. B) calculates a permutation $\sigma_d[i]$ (resp. $\sigma'_d[j]$) representing the reverse of the most likely tidying permutation on $\Psi$ (resp. $\Psi'$) to produce i (resp. j). In other words, with i, $\sigma_d[i]$ realizes :*

$$\max_{\sigma\in\mathfrak{S}_n}\int_x P(i|x)\Psi\circ\sigma_\Psi\circ\sigma^{-1}(x)dx$$

*Then A (resp. B) draws a boolean $b \in \{0,1\}$ (resp. $b'$) and publishes in a random order $(\mu_1, \mu_2) = t^b(\sigma_d[i], \sigma_\Phi)$, (resp. $(\mu'_1, \mu'_2) = t^{b'}(\sigma'_d[j], \sigma_{\Phi'})$) where t represents the transposition of elements in a couple.*

*(we remark that it is possible to publish only* $(\mu_1(I_0),\mu_2(I_0))$ *and* $(\mu'_1(I_0),\mu'_2(I_0))$ *rather than* $(\mu_1,\mu_2)$ *and* $(\mu'_1,\mu'_2)$*, which leads to quantity of bits of* $4n$ *instead of* $4n\log n$*).*

***Step 4 – Synchronization:*** *A (resp.* B*) chooses randomly* $\sigma_A$ *(resp.* $\sigma_B$*) among* $(\mu'_1,\mu'_2)$ *(resp.* $(\mu_1,\mu_2)$*).*

***Step 5 – Advantage Distillation:*** *A calculates* $V_A = \frac{\sigma_\Phi{}^{-1}(x).\sigma_A{}^{-1}(j)}{n}$*, B calculates* $V_B = \frac{\sigma_B{}^{-1}(i).\sigma_{\Phi'}{}^{-1}(y)}{n}$*.* $V_A$ *and* $V_B$ *are then transformed respectively by A and B in binary output thanks to the sampling method:*

$$\widetilde{e_A} = \left\lfloor \frac{(V_A-\rho)\sqrt{nk}}{K} \right\rfloor \bmod 2,\ \widetilde{e_B} = \left\lfloor \frac{(V_B-\rho)\sqrt{nk}}{K} \right\rfloor \bmod 2$$

*(where* $\rho$ *is a translation parameter randomly picked in* $\left[0,\frac{2K}{\sqrt{nk}}\right[$ *at each instance of the protocol). The multiplicative factor* $K$ *is chosen such that:*

$$\frac{1}{\sqrt{nk}} \ll \frac{K}{\sqrt{nk}} \ll \frac{1}{k}$$

*We introduce the following notations corresponding to the canonical form of the opponent strategy to estimate* $V_A$ *(obtained in [9] Proposition 9):*

$$V_\xi(i,j,\sigma_{\xi,A},\sigma_{\xi,B}) \triangleq \frac{2k\left(\left(\sum_{r\in\sigma_{\xi,B}(I_0)} i_r\right)\left(\sum_{r\in\sigma_{\xi,A}(I_0)} j_r\right) + \left(\sum_{r\in\overline{\sigma_{\xi,B}(I_0)}} i_r\right)\left(\sum_{r\in\overline{\sigma_{\xi,A}(I_0)}} j_r\right)\right)}{n^2}$$

$$\widetilde{e_\xi}(\sigma_{\xi,A},\sigma_{\xi,B}) \triangleq \left\lfloor \frac{(V_\xi(i,j,\sigma_{\xi,A},\sigma_{\xi,B})-\rho)\sqrt{nk}}{K} \right\rfloor \bmod 2$$

$$T_m = \left\{\widetilde{e_\xi}(\mu'_1,\mu_1)_m, \widetilde{e_\xi}(\mu'_1,\mu_2)_m, \widetilde{e_\xi}(\mu'_2,\mu_1)_m, \widetilde{e_\xi}(\mu'_2,\mu_2)_m\right\}$$

*where* $\sigma_{\xi,B}$ *is the permutation chosen by* $\xi$ *among* $(\mu_1,\mu_2)$ *as* $\sigma_\Phi$ *and* $\sigma_{\xi,A}$ *is the permutation chosen by* $\xi$ *among* $(\mu'_1,\mu'_2)$ *as* $\sigma_A$*. The partners discard then the instance of the protocol if* $|T_m| \neq 2$*.*

***Step 6:*** *classical Information Reconciliation and Privacy Amplification (IRPA) techniques then lead to get accuracy as close as desired from perfection between estimations of legitimate partners, and knowledge as close as desired from zero by any unlimited opponent, as shown in [4], [11].*

The choices of the parameters $(\alpha,n,k,K,L)$ are theoretically discussed in proof of main Theorem in [9]. They are set to make steps 5 and 6 possible.

The protocol can be heuristically analyzed as follows:

Regarding the legitimate partners:

(1) when $B$ picks $\sigma_B = \sigma_\Phi$ and $A$ picks $\sigma_A = \sigma_{\Phi'}$ (1/4 of cases), the choice of $\sigma_A$ and $\sigma_B$ remain independant from $i,j$, so that $i$ and $j$ remain draws of independent Bernoulli random variables, then allowing to apply Chernoff-style bounds with accuracy $O\left(\frac{1}{\sqrt{nk}}\right)$ ($E\left[\left(V_{A|\sigma_A=\sigma_{\Phi'}} - V_{B|\sigma_B=\sigma_\Phi}\right)^2\right]^{1/2} = O\left(\frac{1}{\sqrt{nk}}\right)$ as shown in [9]).

(2) When $A$ picks $\sigma_A = \sigma'_d[j]$ and/or $B$ picks $\sigma_B = \sigma_d[i]$ ($3/4$ of cases), the Chernoff bound no longer applies and instead, and $V_B$ or $V_A$ become erratic, which will lead to an error probability of $\approx 1/2$.

On the other hand, the discarding condition $|T_m| = 2$ forced at step 5 imposes that the opponent has no better choice than to choose its bit $\widetilde{e_\xi}$ randomly within $\{0,1\}$ with 50% chance. Thus, as soon as we can prove that condition $|T_m| = 2$ and condition (1) above coexist with a fair probability $\not\approx 0$, we have shown that an Advantage is created for the partners over the opponent. This fact is proven in [9] Theorem 1. The main argument of the proof is that, in the estimation $V_\xi(i,j,\sigma_{\xi,A},\sigma_{\xi,B})$ of $V_A$ by opponent, the opponent's choices among $(\sigma_\Phi, \sigma_d[i])$ for $\sigma_{\xi,B}$ and among $(\sigma_{\Phi'}, \sigma'_d[i])$ for $\sigma_{\xi,A}$ are indistinguishable under Deep Random Assumption, and therefore the choice $\left(\sigma_{\xi,A},\sigma_{\xi,B}\right) \neq (\sigma_{\Phi'},\sigma_\Phi)$, which has a fair occurrence probability (3/4), leads to opponent's erratic behavior creating a fair occurrence probability that condition $|T_m| = 2$ and condition (1) above coexist.

Below we explain the role of each step:

The Degradation transformations $x \longmapsto \frac{x}{k}$ and $y \longmapsto \frac{y}{k}$ with $k > 1$ at step 2 are necessary so that $\sigma_d[i](I_0) \neq \sigma_\Phi(I_0)$ (resp. $\sigma'_d[i](I_0) \neq \sigma_{\Phi'}(I_0)$). If $k = 1$, then $i = x$ and $j = y$ and this would cause $\sigma_d[i](I_0) \approx \sigma_\Phi(I_0)$ (resp. $\sigma'_d[i](I_0) \approx \sigma_{\Phi'}(I_0)$).

The Deep Random Generation at step 1 prevents the use of Bayesian inference based on the knowledge of the probability distribution. In particular, if $\Phi, \Psi$ (resp. $\Phi', \Psi'$) were not managed by Deep Randomness, $\xi$ would be able discriminate $\sigma_\Phi$ among $(\mu_1,\mu_2)$ (resp. $\sigma_{\Phi'}$ among $(\mu'_1,\mu'_2)$) by Bayesian inference.

Dispersion step 3 mixes $\sigma_\Phi$ within $(\mu_1,\mu_2)$ with another permutation $\sigma_d[i]$ (and $\sigma_{\Phi'}$ within $(\mu'_1,\mu'_2)$ with another permutation $\sigma'_d[j]$). $i$ is entirely determined by $|i|$ and a permutation, which explains the constraint and transformation applied on $\Psi$ in step 3 to make $\sigma_\Phi$ and $\sigma_d[i]$ indisguishable knowing $i$ (same with $\sigma'_d[j]$, $\sigma_{\Phi'}$, and $j$). $\sigma_d[i]$ is (1) indistinguishable from $\sigma_\Phi$ knowing $\mathfrak{I}_> = \{i, j, (\mu_1,\mu_2), (\mu'_1,\mu'_2)\}$, and (2) causes that $\sigma_d[i]^{-1}(i)$ and $\sigma_\Phi{}^{-1}(i)$ behave very differently due to the fact that $\sigma_d[i]^{-1}(i)$ only depends on $|i|$; indeed we remark that $\forall \mu \in \mathfrak{S}_n\ \sigma_d[\mu(i)]^{-1} \circ \mu = \sigma_d[i]^{-1}$, and therefore, $\sigma_d[i]^{-1}(i)$ is stable by action of $\mathfrak{S}_n$ on $i$. (resp. $\sigma'_d[j]^{-1}(j)$ only depends on $|j|$). That indistinguishability can be expressed under Deep Random assumption by the symmetry $\Sigma_1$ associated to the group $\{\mathrm{Id}, t\}$ ($t$ being the transposition of pairs) applied to $(\mu_1,\mu_2)$ and $(\mu'_1,\mu'_2)$. Associating $\Sigma_1$ with other symmetries presented in the proof of [9] Theorem 1, we manage to show that the best strategy of the opponent under Deep Random assumption to estimate $V_A$ and then as a consequence $\widetilde{e_A}$, are:

$$V_\xi(i,j,\sigma_{\xi,A},\sigma_{\xi,B}) \triangleq \frac{2k\left(\left(\sum_{r\in\sigma_{\xi,B}(I_0)} i_r\right)\left(\sum_{r\in\sigma_{\xi,A}(I_0)} j_r\right) + \left(\sum_{r\in\overline{\sigma_{\xi,B}(I_0)}} i_r\right)\left(\sum_{r\in\overline{\sigma_{\xi,A}(I_0)}} j_r\right)\right)}{n^2}$$

$$\widetilde{e_\xi}(\sigma_{\xi,A}, \sigma_{\xi,B}) \triangleq \left\lfloor \frac{(V_\xi(i,j,\sigma_{\xi,A},\sigma_{\xi,B}) - \rho)\sqrt{nk}}{K} \right\rfloor \bmod 2$$

where $\sigma_{\xi,B}$ is the permutation chosen by $\xi$ among $(\mu_1, \mu_2)$ as $\sigma_\Phi$ and $\sigma_{\xi,A}$ is the permutation chosen by $\xi$ among $(\mu'_1, \mu'_2)$ as $\sigma_A$.

The synchronization step 4 needs that the distributions to have special properties ($\in \zeta(\alpha)$) in order to efficiently play their role. It is efficient in $1/4$ of cases (when $B$ picks $\sigma_B = \sigma_\Phi$ and $A$ picks $\sigma_A = \sigma_{\Phi'}$, which we will call 'favorable cases'). If $\sigma_\Phi$ and $\sigma_{\Phi'}$ were chosen randomly by $A$ and $B$ instead of being the respective tidying permutations of $\Phi$ and $\Phi'$, then $\xi$ could, even without knowing $\Phi$ and $\Phi'$, use $V_{\xi,1}(i,j) \triangleq \frac{k|i||j|}{n^2}$ in order to estimate say $V_A$ with accuracy $O\left(\frac{1}{\sqrt{nk}}\right)$, in which case $\widetilde{e_\xi} = \left\lfloor \frac{(V_{\xi,1}-\rho)\sqrt{nk}}{K} \right\rfloor$ would be as close to $\widetilde{e_A}$ as $\widetilde{e_B}$ in favorable cases.

The step 5 is called Advantage Distillation because at this step, thanks to the discarding condition $|T_m| = 2$, we have managed to create, under Deep Random Assumption, a protocol in which the error rate for the legitimate receiver is strictly lower than the error rate for the opponent.

We remark that the condition $|T_m| = 2$ plays also another role which is to protect against the unexpected predictability effects of using an error correcting code at Reconciliation phase. Indeed, unlike QKD protocols (protected by no cloning theorem) or Maurer's satellite protocol presented in [5], in our case the opponent can try the various possible combinations of choice for $\sigma_{\xi,A}$ and $\sigma_{\xi,B}$, and compare them in terms of matching a code word. To illustrate the issue, let's consider the non-optimal 'bit repeating' error correction method described by Maurer in [5]: the codeword $v_A$ chosen by $A$ can only be $(0,0,\dots,0)_L$ or $(1,1,\dots,1)_L$ depending on $e_A = 0$ or $e_A = 1$. $B$ could then publicly discards all decoded sequence $v_B$ that is not $(0,0,\dots,0)_L$ or $(1,1,\dots,1)_L$ and obviously decodes accordingly $e_B = 0$ if $|v_B| = 0$, and $e_B = 1$ if $|v_B| = L$. In such scenario, if, for one of the instance $m$ within the sequence, we have $\sigma_A = \sigma_{\Phi'}$ and $T_m = (0,0,0,0)$ (or $T_m = (1,1,1,1)$), then it is clear that the code word is $(0,0,\dots,0)_L$ (or $(1,1,\dots,1)_L$). This example shows how, even with Deep Randomness indistinguishability applied locally to each instance, the opponent can still take advantage of an imprudent implementation of the error correcting method to break the secrecy. The condition $|T_m| = 2$ also have the advantage to avoid such predictability flaw.

## II. The "bit reuse and recombine" optimization technique

The protocol described in Section I requires to exchange $6n$ bits $(i, j, \mu_0(I_0), \mu_1(I_0), \mu'_0(I_0), \mu'_1(I_0))$ in order to distill $\widetilde{e_A}$ and $\widetilde{e_B}$. This is extremely costly, considering that $n$ should be sufficiently large in order to let the law of large numbers produce its effects with the Bernoulli variables $V_A$ and $V_B$ in favorable cases.

The technique that we present in this section makes this multiplication factor to vanish by reusing draws without decreasing the entropy of the output digit flow. It thus represents a dramatic improvement of performance.

Here is how it works.

When the protocol is used to transmit a message $M$, the transmission is sequenced in successive blocks of $n$ instances of the steps 1 to 4 of the protocol. Each block enables to distill $n^2$bits $\widetilde{e_A}$ and $\widetilde{e_B}$ with a reuse and recombine technique that is detailed below. Then Information Reconciliation and Privacy Amplification (IRPA) can be applied on those $n^2$bits flows to distill fully secure $e_A$ and $e_B$ bit flows.

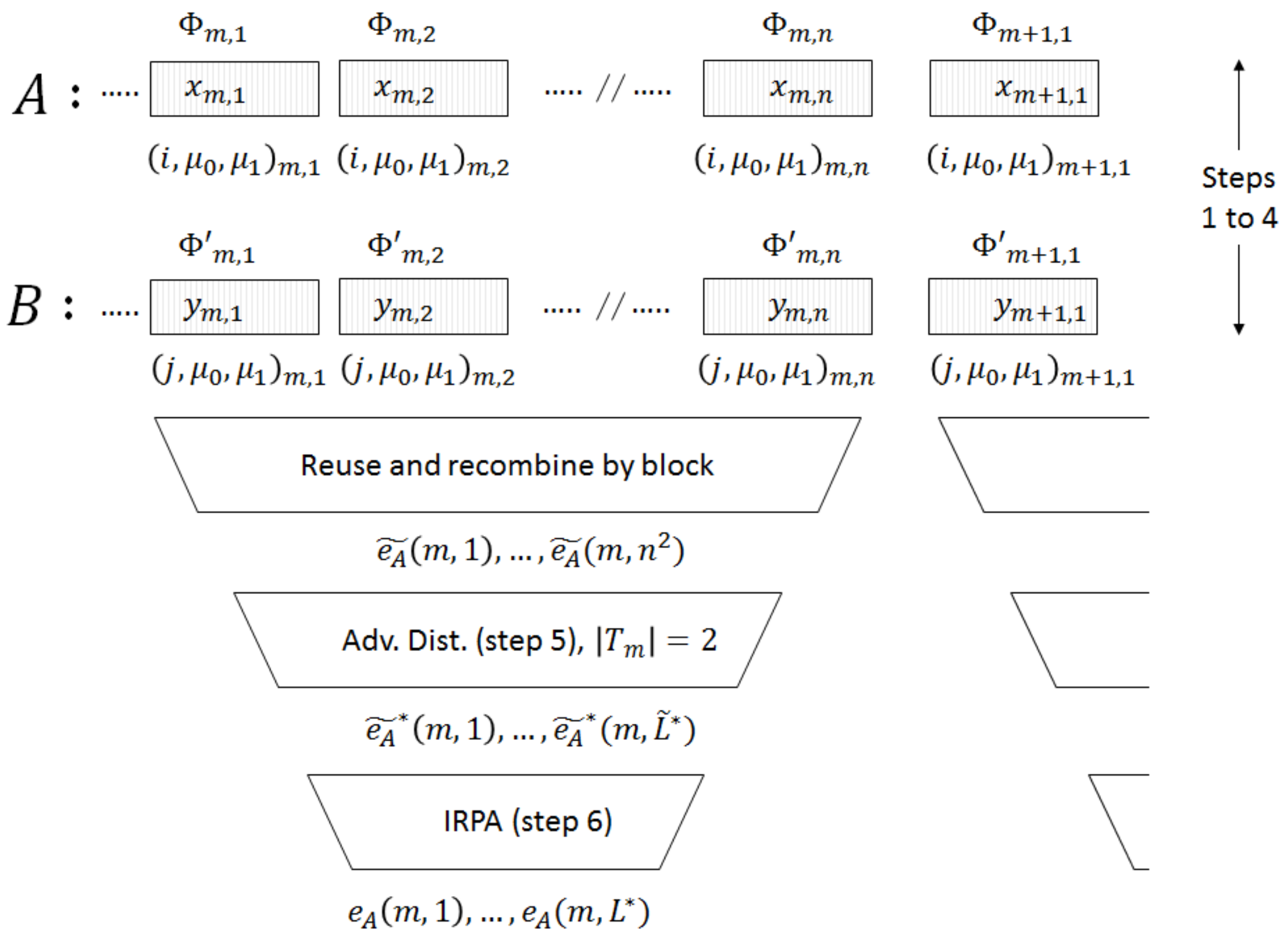

Fig.1

We denote by $m$ the sequence number of a transmission block, and $q \in \mathbb{N}_n^*$ the instance number, and $p \in \mathbb{N}_{n^2}^*$ the index of the output $\widetilde{e_A}$ (resp. $\widetilde{e_B}$) digit. Then, the recombination uses the following variables:

$$V_A(m,p) \triangleq \frac{\sigma_{\Phi_{m,\left\lfloor\frac{p-1}{n}\right\rfloor+1}}^{-1}\left(x_{m,\left\lfloor\frac{p-1}{n}\right\rfloor+1}\right)\cdot {\mu'}_{b_{m,p},m,((p-1) \bmod n)+1}^{-1}(j_{m,((p-1) \bmod n)+1})}{n}$$

$$V_B(m,p) \triangleq \frac{\sigma_{\Phi'_{m,((p-1) \bmod n)+1}}^{-1}(y_{m,((p-1) \bmod n)+1})\cdot \mu_{b'_{m,p},m,\left\lfloor\frac{p-1}{n}\right\rfloor+1}^{-1}\left({i'}_{m,\left\lfloor\frac{p-1}{n}\right\rfloor+1}\right)}{n}$$

$$\widetilde{e_A}(m,p) \triangleq \left\lfloor\frac{(V_A(m,p)-\rho_{m,p})\sqrt{nk}}{K}\right\rfloor \bmod 2$$

$$\widetilde{e_B}(m,p) \triangleq \left\lfloor\frac{(V_B(m,p)-\rho_{m,p})\sqrt{nk}}{K}\right\rfloor \bmod 2$$

We have to justify that (conservation of Entropy):

$$H\left(\{\widetilde{e_A}(m,p)\}_{p\in\mathbb{N}_{n^2}^*}\right) = H\left(\{\widetilde{e_B}(m,p)\}_{p\in\mathbb{N}_{n^2}^*}\right) \approx n^2$$

Let's take $\{\widetilde{e_A}(m,p)\}_{p\in\mathbb{N}_{n^2}^*}$ (by symmetry, the result is the same for $\{\widetilde{e_B}(m,p)\}_{p\in\mathbb{N}_{n^2}^*}$). One can view

$$\{V_A(m,p)\}_{p\in\mathbb{N}^*_{n^2}}$$

as a linear system of size $n^2$. Let's consider the matrix $J_m$ whose line vectors are the ${\mu'_{b_{m,q}}}^{-1}(j_{m,q})$. The linear system above can be written as block matrix

$$\begin{pmatrix} J_m & & & 0 \\ & J_m & & \\ & & \ddots & \\ 0 & & & J_m \end{pmatrix}^t \left({\sigma_{\Phi'_{m,1}}}^{-1}(y_{m,1}), \dots, {\sigma_{\Phi'_{m,n}}}^{-1}(y_{m,n})\right)$$

For each $J_m$ block, we can apply the result of Proposition AII-2 in Appendix II, with the value of binary random variable parameter being:

$$\theta = \frac{1}{nk}\int_{[0,1]^n} |x|\Phi_{m,r}(x)dx$$

and the sampling parameter $q$ being:

$$\delta = \frac{K\sqrt{n}}{\sqrt{k}}$$

We then have $\delta^n\sqrt{n!\,\theta^n(1-\theta)^{n-1}(n\theta-\theta+1)} \gg 1$, and thus $H\left(\{\widetilde{e_A}(m,p)\}_{p\in\mathbb{N}^*_{n^2}}\right) \approx n^2$. We note as well ([15], [16]) that the probability of a random $\{0,1\}$ matrix is singular is $O\left(\frac{1}{\sqrt{n}}\right)$.

With this technique, $A$ (resp. $B$) generates only $n$ pairs $(\mu_0(I_0), \mu_1(I_0))$ rather than $n^2$ pairs in order to obtain the $n^2$ distilled digits $\{\widetilde{e_A}(m,p)\}_{p\in\mathbb{N}^*_{n^2}}$.

Lastly, the fact that the lines of $J_m$ are generated by a distribution in $\zeta(\alpha)$ rather than a uniform distribution over $\{0,1\}^n$ does not jeopardize its entropy, and thus keep the assumption that is nonsingular fairly reasonable. As an example of the above statement, let's determine $H(x)_{\Phi_0}$ where $\Phi_0$ is the distribution given by:

$$P\left(\sum_{s=1}^{n/2} x_s = t\right) = P\left(\sum_{s=n/2+1}^{n} x_s = t'\right) = \frac{2}{n}, \forall t, t' \in \{0, \dots, n/2\}$$

Its quadratic matrix is:

$$\begin{pmatrix} 1/2 & \cdots & 1/3 & 1/4 & \cdots & 1/4 \\ \vdots & \ddots & \vdots & \vdots & & \vdots \\ 1/3 & \cdots & 1/2 & 1/4 & \cdots & 1/4 \\ 1/4 & \cdots & 1/4 & 1/2 & \cdots & 1/3 \\ \vdots & & \vdots & \vdots & \ddots & \vdots \\ 1/4 & \cdots & 1/4 & 1/3 & \cdots & 1/2 \end{pmatrix}$$

and it verifies (with notations introduced in [9]):

$$\left\|M_{\Phi_0} - \overline{M_{\Phi_0}}\right\|_c = \frac{1}{12} + O\left(\frac{1}{n}\right)$$

We show in Appendix I that $H(x)_{\Phi_0} = \frac{n}{2} - O(\log n)$, which suggests that the probability that $J_m$ is singular remains small even with a distribution in $\zeta(\alpha)$.

**Simulations**

Simulations for the protocol with 'Reuse and Recombine' optimization technique are presented in Appendix III.

**Memory vs Bandwidth compromise**

It is clear that this technique dramatically improves the bandwidth consumption, but the price to pay is a dramatic increase of the memory needed by each partner to perform the optimized protocol. Indeed they have to create and maintain a memory structure of $n^2$ bits during the duration of the protocol.

On the other hand, it is possible to take an intermediate stance, where the partners only memorize $w < n$ vectors instead of $n$. The reuse and recombine method then follows the same logic, the memory structure are $nw$ bits large instead of $n^2$, and the bandwidth reduction factor will be $w$ instead of $n$.

The parameter $w$ can be chosen by a partner as his maximum possibility with regards to his memory capacity, and the corresponding partners can then negotiate $w = \min(w_A, w_B)$ during an initial phase of the protocol. A Server-Server communication in a rare bandwidth context will use a maximum $w = n$, when a Client-Server will use a smaller $w$ depending on the Client capacity.

**Combination of error correcting codes**

The bit repetition error correcting method used in [9] is clearly not optimal in terms bandwidth consumption, but it has the power to lower error rate starting from a very high value (close to the maximum 50%). Typically, in the simulations of Appendix III, we use basic bit repetition codes and we satisfy ourselves only by lowering the partners' error rate below 30%, which corresponds to a BSC error parameter below 15%. It is established in [13] that, when the BSC error parameter is below 15%, it is possible to use Cascade, a more efficient error correcting protocol than the bit repetition one. Cascade has the drawback to be highly interactive, but it is close to optimal efficiency in terms of bandwidth consumption. It has also been proposed [18] to use optimized versions for BSC of LDPC codes initially introduced by Gallager [17], in order to avoid the interactive complexity of Cascade. Both Cascade and optimized LDPC for BSC work for a range of protocols including QKD and Maurer satellite scenario, and they are also applicable to the Deep Random Secrecy protocols in their Reconciliation phase.

Therefore, it is recommended to adapt the protocol parameters in such a way that the repetition code at step 6 is used with the smallest possible code length needed to drive the error rate below the threshold where one of the above Reconciliation protocols become efficient, in order to reach an acceptable error rate.

## III. The 'non-contributive avoidance' optimization technique

In this section, we present another, complementary, optimization technique, that avoids non-contributive instances. At step 4, $A$ (resp. $B$) choose randomly $\sigma_A$ *(resp.* $\sigma_B$*)* among $(\mu'_1, \mu'_2)$ (resp.

$(\mu_1, \mu_2)$). Those choices lead to couples of alternative possible digit $\left(\widetilde{e_A}(\mu'_1), \widetilde{e_A}(\mu'_2)\right)$ for $A$ and $(\widetilde{e_B}(\mu_1), \widetilde{e_B}(\mu_2))$ for $B$.

It is clear that if, for a given instance of the protocol, $\widetilde{e_A}(\mu'_1) \neq \widetilde{e_A}(\mu'_2)$ or $\widetilde{e_B}(\mu_1) \neq \widetilde{e_B}(\mu_2)$, then that instance is equivalent to a situation where one of the legitimate partner picks its digit 50% randomly, which means that such instance is useless for the purpose of reaching mutual agreement with lowest possible error rate. We call such instances 'non-contributive'. It is obvious that reducing the ratio of non-contributive instances also decreases the error rate of the legitimate receiver.

One can use the parameters $K$ and $\rho$ to avoid non-contributive instances. Here is how this can be achieved.

We first remark that, if the following condition is satisfied:

$$|V_B(\mu_1) - V_B(\mu_2)| = \frac{2lK}{\sqrt{nk}}$$

then $\widetilde{e_B}(\mu_1) = \widetilde{e_B}(\mu_2)$, $\forall l \in \mathbb{N}$. Therefore, $B$ can adapt the value of $K$ as soon as it gets knowledge of $\{i, (\mu_1, \mu_2)\}$ as follows:

$$l \triangleq \left\lceil \frac{|V_B(\mu_1) - V_B(\mu_2)|}{2K} \right\rceil$$

$$K \to \frac{|V_B(\mu_1) - V_B(\mu_2)|}{2l}$$

With that first adaptation of parameter $K$, $B$ can ensure that $(\widetilde{e_B}(\mu_1), \widetilde{e_B}(\mu_2))$ is contributive. One can also remark that the above adaptation is not affected by a variation of the translation parameter $\rho$. That will enable $A$ to, in turn, adapt the parameter $\rho$ in order to ensure that $\left(\widetilde{e_A}(\mu'_1), \widetilde{e_A}(\mu'_2)\right)$ is contributive, without jeopardizing the adaptation of $B$. To that extend, $A$ will adapt $\rho$ such that

$$\widetilde{e_A}(\mu'_1) \triangleq \left\lfloor \frac{\left(V_A(\mu'_1) - \rho\right)\sqrt{nk}}{K} \right\rfloor \bmod 2 = \left\lfloor \frac{\left(V_A(\mu'_2) - \rho\right)\sqrt{nk}}{K} \right\rfloor \bmod 2 \triangleq \widetilde{e_A}(\mu'_2)$$

which is almost always possible (except when $|V_A(\mu'_1) - V_A(\mu'_2)| = \frac{2lK}{\sqrt{nk}}$).

The protocol can be adapted as follows to benefit from 'non-contributive avoidance' optimization technique (we omit in the following description to modify the protocol also with the 'bit reuse and recombine' optimization technique, but it is obvious that they can co-exist without any difficulty because they operate on distinct aspects of the protocol):

A step 4' can be added between Step 4 and Step 5

***Step 4' – Adaptation of $K, \rho$:*** *$B$ computes the new value of $K$ based on initial value of $K$ and on $\{i, (\mu_1, \mu_2)\}$. $B$ publishes the new value of $K$. $A$ computes the new value of $\rho$ based on the new value of $K$ and on $\{j, (\mu'_1, \mu'_2)\}$. $A$ publishes the new value of $\rho$.*

When combined with the 'bit reuse and recombine' optimization technique, the values of $K, \rho$, being different for each instance, will thus be noted $K_{m,p}, \rho_{m,p}$.

It should be noted that publishing the new value of $K$ does not give useful information to the opponent. Indeed, it is an immediate consequence of [9] Lemma 1 that $|\mu_1(i) \cap \mu_2(i)| \gg K/\sqrt{nk}$, and therefore the knowledge of condition $|V_B(\mu_1) - V_B(\mu_2)| = \frac{2lK}{\sqrt{nk}}$ does not give sufficiently precise information about $y$ to be able to determine $\widetilde{e_B}$.

Similarly, the publication of $\rho$ does not neither give useful information to the opponent regarding $x$.

## IV. Extended protocol against active opponent

We have introduced in [14] how the protocol reminded in Section I can be extended to include an authentication scheme, enabling the legitimate partners to resist to active unlimited MITM. In this section, we will (i) remind the results obtained in [14], (ii) show how the extension copes with the "bit reuse and recombine" optimization technique, and the 'non-contributive avoidance' optimization technique seen in former sections, and (iii) discuss some implementation aspects.

The purpose of the extended protocol is still to generate a common secret bit string $S$ between the legitimate partners, but this time in the perspective of an active unlimited MITM.

Without the authentication extension, the protocol can be represented by the diagram below with the "bit reuse and recombine" optimization technique:

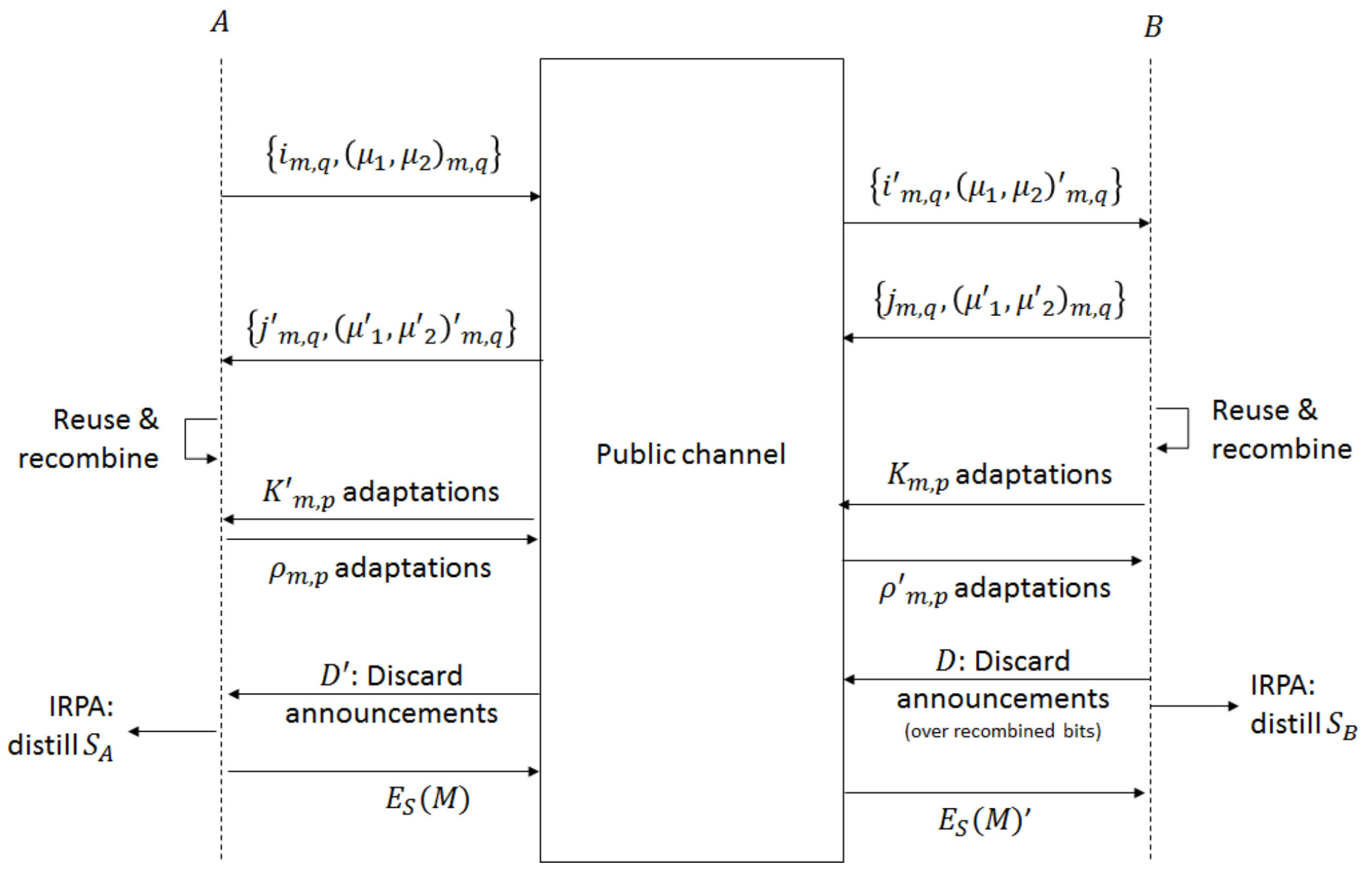

Fig. 2

The purpose of generating $S$ is of course to transmit a secret message $M$ from $A$ to $B$. The encryption function $E_S(M)$ can be simply $S \oplus M$ if one want the highest security offered by one time pad, or another symmetric key encryption scheme working with a shared secret key $S$. While we have shown that a passive opponent cannot gain knowledge of $S$, it is obvious that an active opponent can insert himself in the communication, first playing the role of $B$ with $A$, which enables him to collect $M$, and then, optionally, to play the role of $A$ with $B$ to transmit $M$, or even a different message $M'$ to $B$.

We will see now how to prevent this MITM attack with an extended version of the protocol $\mathcal{P}$.

**Reminder of the security model**

In our 'active opponent scenario', $\xi$ still has unlimited computation and storage power, but is also capable to (i) delete messages from the public channel, (ii) introduce new messages in the public channel at destination of either $A$ or $B$; (i)+(ii) is also equivalent to the capacity to modify a message transiting from $A$ to $B$ or from $B$ to $A$.

$A$ and $B$ are equipped with a DRG, and also with a private 'wallet' containing a shared authentication secret $s$, that can be updated at any time by the wallet holder. It is assumed as a pre-condition to the protocol that $A$ and $B$ have initially the same value $s_0$ in their wallet.

It is assumed that the opponent $\xi$ has no access to (i) the content of a private wallet, (ii) anything computed or stored within a DRG.

The goal of the extended protocol is to continue to ensure that, $\forall \varepsilon, \varepsilon' > 0$

*(i)* $P(e_A \neq e_B) \leq \varepsilon$

*(ii)* $\left| P(e_\xi = e_B) - \frac{1}{2} \right| \leq \varepsilon'$

even with such an active opponent.

**Description of the extended protocol $\mathcal{P}^*$**

As formerly exposed in [14], the extended protocol that we propose is possible only because of a specific non-reversibility property of $\mathcal{P}$. Indeed it is shown in [9] (Lemma3) that, when the public information $(i,j)$ is given, with the assumption that the distributions are synchronized, then it is impossible to reversely determine $x$ or $y$ with an accuracy equal to the one of a partner's estimation. $\mathcal{P}^*$ is original because, unlike any other protocol that does not benefit from the non-reversibility property of Deep Random Secrecy, it resists to opponents that are both unlimited and active MITM.

The extension presented in [14] consists in adding a mutual verification phase after the bit string $S$ has been generated, and before transmitting the secret message $M$. $M$ having a length bounded by $|M| \leq L_M$, we denote by $N$ the number of blocks to be sent to distill a message of length $L_M$; typically $N = \lceil L_M / L^* \rceil$ (using notations of Fig. 1). This verification is performed by:

(i) dividing the shared secret into 2 independent piece $(s_A, s_B)$ (with $H(s_A) = H(s_B) = H(s)/2$)

(ii) sending from $A$ to $B$ (resp. $B$ to $A$) a verification code

$$c_A = h\left(s_A, \{i_{m,q}, (\mu_1, \mu_2)_{m,q}\}_{m \leq N, q \in \mathbb{N}_n^*}, \{j'_{m,q}, (\mu'_1, \mu'_2)'_{m,q}\}_{m \leq N, q \in \mathbb{N}_n^*}\right)$$

(resp.)

$$c_B = h\left(s_B, \{i'_{m,q}, (\mu_1, \mu_2)'_{m,q}\}_{m \leq N, q \in \mathbb{N}_n^*}, \{j_{m,q}, (\mu'_1, \mu'_2)_{m,q}\}_{m \leq N, q \in \mathbb{N}_n^*}\right)$$

(the characteristics of the verification function $h$ will be discussed hereafter)

(iii) verifying by $B$ (resp. by $A$) that the local computation

$$v_A = h\left(s_A, \{i'_{m,q}, (\mu_1, \mu_2)'_{m,q}\}_{m \leq N, q \in \mathbb{N}_n^*}, \{j_{m,q}, (\mu'_1, \mu'_2)_{m,q}\}_{m \leq N, q \in \mathbb{N}_n^*}\right)$$

(resp.)

$$v_B = h\left(s_B, \{i_{m,q}, (\mu_1, \mu_2)_{m,q}\}_{m \leq N, q \in \mathbb{N}_n^*}, \{j'_{m,q}, (\mu'_1, \mu'_2)'_{m,q}\}_{m \leq N, q \in \mathbb{N}_n^*}\right)$$

is equal to the received code $(c'_A = v_A)$ (resp. $(c'_B = v_B)$

(iv) using part of the entropy of $S$ to renew (sending $R(s')$) a new value of the shared authentication secret $s'$ for a next round

(v) storing by $A$ (resp. $B$) the new value of shared authentication secret $s'$ in its private wallet.

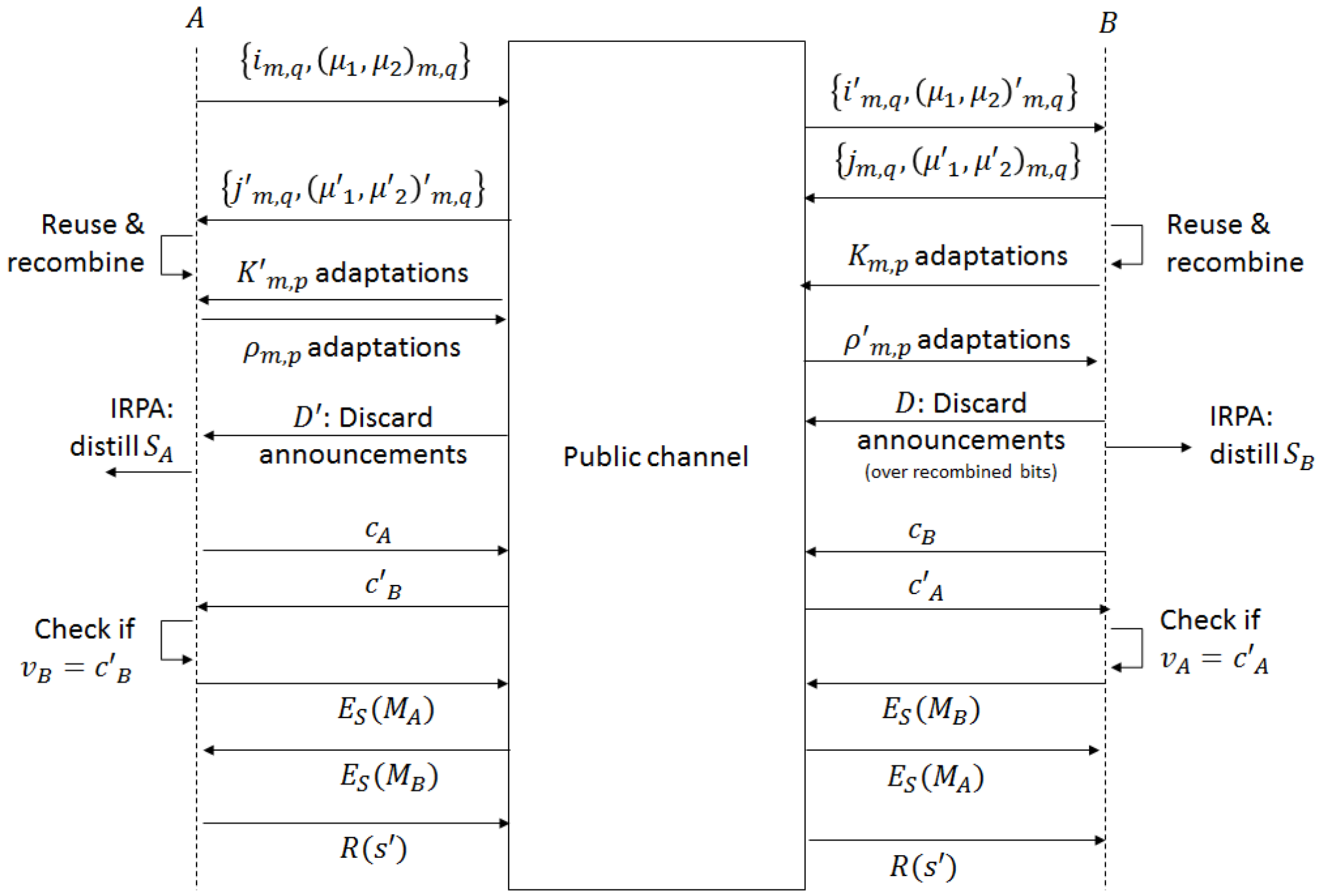

Fig. 3

Assuming that $E_S(M) = S \oplus M$ (secret messages can be sent in both sense, which is why the diagram above represents $E_S(M_A)$ and $E_S(M_B)$), the common shared bit string must have length:

$$|S| \geq H(s) + L_M$$

in order to be able to fully renew the shared authentication secret $s$ after each round.

It should be noted that the above described example presents a mutual authentication. It is of course possible to execute the protocol with a one way authentication only if for instance, say $B$, is a public resource, in which base only $B$ needs to authenticate to $A$.

**Design considerations for security**

We have shown in [14] that the authentication scheme is secure if the verification function $h$ satisfies the following sufficient (but not necessary) properties:

Property 1:

$$\textit{if } \{i_{m,q}, (\mu_1, \mu_2)_{m,q}\}_{m \leq N, q \in \mathbb{N}_n^*}, \{j'_{m,q}, (\mu'_1, \mu'_2)'_{m,q}\}_{m \leq N, q \in \mathbb{N}_n^*} \textit{ are known, then}$$

$$\# \left\{ s \,\middle|\, h\left(s, \{i_{m,q}, (\mu_1, \mu_2)_{m,q}\}_{m \leq N, q \in \mathbb{N}_n^*}, \{j'_{m,q}, (\mu'_1, \mu'_2)'_{m,q}\}_{m \leq N, q \in \mathbb{N}_n^*}\right) = h\left(s_A, \{i_{m,q}, (\mu_1, \mu_2)_{m,q}\}_{m \leq N, q \in \mathbb{N}_n^*}, \{j'_{m,q}, (\mu'_1, \mu'_2)'_{m,q}\}_{m \leq N, q \in \mathbb{N}_n^*}\right) \right\} \gg 1$$

Property 2:

$$if\ \{i_{m,q},(\mu_1,\mu_2)_{m,q}\}_{m\le N,q\in\mathbb{N}_n^*},\{j'_{m,q},(\mu'_1,\mu'_2)'_{m,q}\}_{m\le N,q\in\mathbb{N}_n^*}\ are\ known,\ and\ s\neq s'$$

$$P\left(h\left(s,\{i_{m,q},(\mu_1,\mu_2)_{m,q}\}_{m\le N,q\in\mathbb{N}_n^*},\{j'_{m,q},(\mu'_1,\mu'_2)'_{m,q}\}_{m\le N,q\in\mathbb{N}_n^*}\right)\right.$$
$$\left.= h\left(s',\{i_{m,q},(\mu_1,\mu_2)_{m,q}\}_{m\le N,q\in\mathbb{N}_n^*},\{j'_{m,q},(\mu'_1,\mu'_2)'_{m,q}\}_{m\le N,q\in\mathbb{N}_n^*}\right)\right)=o_{H(s)}(1)$$

Property 3:

*There does not exist any non-trivial transformation $T$ applying on the space $\{i'_{m,q},(\mu_1,\mu_2)'_{m,q}\}_{m\le N,q\in\mathbb{N}_n^*},\{j'_{m,q},(\mu'_1,\mu'_2)'_{m,q}\}_{m\le N,q\in\mathbb{N}_n^*}$ such that*

$$\forall s,\quad h\left(s,\{i_{m,q},(\mu_1,\mu_2)_{m,q}\}_{m\le N,q\in\mathbb{N}_n^*},T\left(\{j'_{m,q},(\mu'_1,\mu'_2)'_{m,q}\}_{m\le N,q\in\mathbb{N}_n^*}\right)\right)$$
$$= h\left(s,T(\{i'_{m,q},(\mu_1,\mu_2)'_{m,q}\}_{m\le N,q\in\mathbb{N}_n^*}),\{j_{m,q},(\mu'_1,\mu'_2)_{m,q}\}_{m\le N,q\in\mathbb{N}_n^*}\right)$$

We propose to use:

$$h_{a,b}(x)\triangleq\big((ax+b)\ \mathrm{mod}\ p\big)\ \mathrm{mod}\ 2^{\frac{H(s_A)}{2}}$$

$$h\left(s,\{i_{m,q},(\mu_1,\mu_2)_{m,q}\}_{m\le N,q\in\mathbb{N}_n^*},\{j_{m,q},(\mu'_1,\mu'_2)_{m,q}\}_{m\le N,q\in\mathbb{N}_n^*}\right)$$
$$\triangleq\left(\bigoplus_{l\in\{(m,q)|m\le N,q\in\mathbb{N}_n^*\}}\left(h_{a_l,b_l}(i_l+s\overline{\iota_l})\right)\right)$$
$$\oplus\left(\bigoplus_{l\in\{(m,q)|m\le N,q\in\mathbb{N}_n^*\}}\left(h_{a'_l,b'_l}(j_l+s\overline{j_l})\right)\right)$$
$$\oplus\left(\bigoplus_{l\in\{(m,q)|m\le N,q\in\mathbb{N}_n^*\}}\left(h_{c_l,d_l}({\mu_1}_l+s\overline{{\mu_1}_l})\right)\right)$$
$$\oplus\left(\bigoplus_{l\in\{(m,q)|m\le N,q\in\mathbb{N}_n^*\}}\left(h_{c'_l,d'_l}({\mu_1}'_l+s\overline{{\mu_1}'_l})\right)\right)$$
$$\oplus\left(\bigoplus_{l\in\{(m,q)|m\le N,q\in\mathbb{N}_n^*\}}\left(h_{e_l,f_l}({\mu_2}_l+s\overline{{\mu_2}_l})\right)\right)$$
$$\oplus\left(\bigoplus_{l\in\{(m,q)|m\le N,q\in\mathbb{N}_n^*\}}\left(h_{e'_l,f'_l}({\mu_2}'_l+s\overline{{\mu_2}'_l})\right)\right)$$

where $p$ is a prime superior to $2^{\frac{H(s)}{2}}$, and $\{a_l,b_l,c_l,d_l,e_l,f_l,a'_l,b'_l,c'_l,d'_l,e'_l,f'_l\}$ are random non-zero integers, fixed as parameters of the protocol (not negotiated); they can typically be the successive outputs of a congruential generator to avoid to store all of them in memory. The permutation ${\mu_1}_l$ is written above with its canonical binary form to be able to compute it with arithmetic. $i_l+s\overline{\iota_l}$ is a non-linear form on $i_l$ also involving $s$ for each $i_l,{\mu_1}_l,{\mu_2}_l,j_l,{\mu'_1}_l,{\mu'_2}_l$. $\{h_{a,b}(x)\}$ is a well known universal hashing class of functions, thus benefiting from the good probabilistic properties of such class.

The modularity structure of $h$ in terms of using blocks of the form $\left(h_{a_l,b_l}(i_l+s\overline{\iota_l})\right)$ and composing them by simple $\oplus$ is chosen on purpose; when the sets $\{i_l,(\mu_1,\mu_2)_l\}$ are reused in our reuse and recombine technique, the associated block can be reused without re-computation of the arithmetical operations.

Eventually, note that the computation of $h$ is obviously highly sensitive to errors in the transmission of the $\{i_l, (\mu_1, \mu_2)_l, j_l, (\mu'_1, \mu'_2)_l\}$, and therefore it is recommended to associate each $\{i_l, (\mu_1, \mu_2)_l\}$ with an error control checksum.

The size parameters $H(s), L_M, L$ should also be fixed as parameters of the protocol and not negotiated, in order to avoid that they become also vulnerable to MITM attacks. The adapted parameters $\{K_{m,p}, \rho_{m,p}\}$ can be let outside of the computation of $h$ because its modification cannot do more for the opponent than increasing the error rate between the legitimate partners.

## V. Implementation considerations

An implementation of Deep Random secrecy for a communicating entity (Partner A) can typically comprise 3 modules, as shown in the diagram below:

- A Deep Random Generator
- A Wallet
- A Communication module

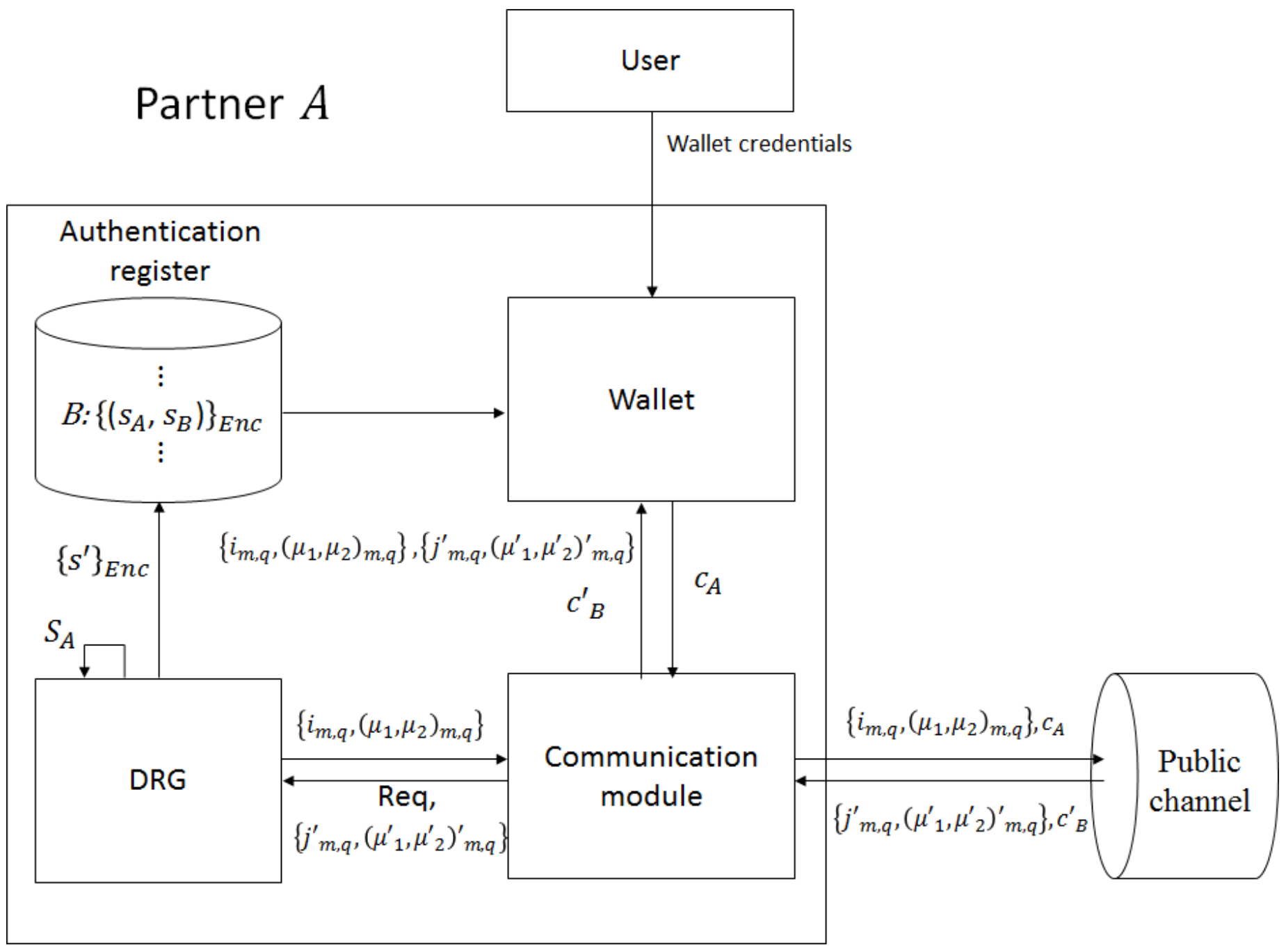

Fig. 4

### The Deep Random Generator (DRG)

The Deep Random Generator can be implemented by using the techniques presented in [12], and in particular the 'partition summing' method to compute the synchronization permutation $\sigma_{\Phi_s}$ for each new distribution $\Phi_s$ generated.

The DRG is supported by a recursive generation process as presented in [12], and should not deliver any usable output before it has reached its 'maturity period' as defined in [12]. Once the maturity period has been reached, one may consider the compromise between entropy of the DRG and computation cost. Indeed, the implementation that maximizes DRG's entropy is to continue running the recursive process without interruption ('continuous' implementation strategy), but it is at the cost of a significant burden on the computation power. Another possible implementation is to stop the recursive process once the DRG has reached maturity, and from that moment to iterate only at the time when the generation of a new vector $x$ is requested. That 'on demand' strategy of implementation respects the minimal entropy requirement together with being much less greedy in computation power. Devices that have limited computation power (Clients) should adopt the 'on demand' implementation strategy. Some powerful entities (DRG servers) may be reserved or dedicated to generate high entropy Deep Randomness, and can then service other entities to "upgrade" their DRG internal state with new possible distributions, by using the combination process described in [12].

It is recommended that a DRG is implemented in a tamper resistant manner, so that the internal state of the generator is kept safe from memory investigation. The scope of the DRG treatments implemented in the tamper resistant environment depends on the tamper resistant memory capacity available. In the ideal, only the public information would be processed outside of the DRG, as suggested in the above diagram.

The Dispersion Step involves the generation of $\sigma_d[i]$. Instead of trying to compute

$$\max_{\sigma \in \mathfrak{S}_n} \int_x P(i|x)\Psi \circ \sigma_\Psi \circ \sigma^{-1}(x)dx$$

it is easier to simply generate an $x'$ out of $\Psi \circ \sigma_\Psi$ and then to search for $\sigma_d[i]$ that realizes:

$$\max_{\sigma \in \mathfrak{S}_n} P(i|\sigma(x'))$$

**The Authentication Wallet**

The authentication Wallet manages the one way or mutual authentication between $A$ and any partner $B_t$. The Wallet is associated with a directory called "Authentication register" where, for each registered partner $B_t$, the current value of the authentication shared secret $(s_A, s_B)_t$ is securely stored (typically encrypted with a master key managed by the Wallet's user and activated by some Wallet credentials as generically suggested in the diagram above). For one way registered partner (authenticated public resources), the authentication shared secret is half empty $(\cdot, s_B)_t$.

The initialization of authentication shared secret between $A$ and $B_t$ happens during a 'registration phase' in which the partners are not yet capable to authenticate each other by mean of the protocol. Thus, some other credential should be exchanged between $A$ and $B_t$ during the 'registration phase' in order for them to trust each other. Those 'identification' credentials and the initial seed of their authentication shared secret should be exchanged through an unauthenticated instance of the protocol in order to protect them in confidentiality and integrity.

Note that there may happen a loss of synchronization of the authentication shared secret between a pair of partners due to network connection failures. Therefore, in order to make the protocol resistant to such loss of synchronization, it should be allowed to reuse the authentication shared secret a limited number of times (which however should be taken into account in the conditional entropy of $c_A$ and $c_B$). Beyond that limit, the shared secret should be considered as compromised and the partners should go again through a registration phase.

**Advanced protocols**

DR-TLS:

The extended protocol presented in this paper supplies primitives to exchange keys, and / or perform a one way or mutually authenticated connection between 2 partners. It is therefore natural to question how such protocol could candidate to replace a popular protocol such as TLS in a post quantum era.

Where our protocol is conceptually even stronger than TLS because it prevents by design MITM attacks, it does not offer the convenience of Public Key infrastructure and signature to easily 'forward' the trust from a root authority to any end-user trusting that authority to authenticate its partner. The concept of Public Key cryptography is, by nature, excluded from our security model because the opponent is supposed to have unlimited computation capacity. However, the issuance by a trusted authority of a digital certificate to one of the partner (single TLS authentication) or to both of the partners (TLS mutual authentication) can be replaced by an interactive confirmation by a trusted authority, supporting the registration phase:

Suppose Partner $A$ and Partner $B$ want to enter into registration phase and avoid MITM attacks. Both are already registered with a trusted authority $C$. Here is how they can proceed:

1) $A$ informs $C$ that it wants to register with $B$, and exchange a session key $s_A$ with $C$, by using the mutually authenticated key exchange primitive.
2) $C$ requests $B$ to exchange a session key $s_B$, by using the mutually authenticated key exchange primitive.
3) $A$ and $B$ use the unauthenticated key exchange primitive to exchange (i) an initial seed of their authentication shared secret, (ii) the values of $c_A$ and $c_B$ as defined in the extended protocol, sent respectively by $A$ and $B$.
4) $C$ sends $s_B$ to $A$ (still through their authenticated channel) which enables $A$ to compute $v_B$ and thus to confirm the identity of $B$. $B$ can do the same by receiving $s_A$ from $C$.

Such method can be refined in many ways, like for instance by using Kerberos-like ticketing model in order to avoid that $C$ is systematically requested by both partners willing to register to each other.

DR-SMTP:

One of the interest of Deep Random Secrecy compared to Quantum Key Distribution (QKD), besides obviously to be fully implementable with classical computing and communication resources, is that Deep Random Secrecy supports asynchronous protocols, due to the fact that the public information exchanged along the protocol can be store, duplicated, forwarded, resent at will without jeopardizing the security or the functioning of the protocol. This is of course not true with quantum information that cannot be duplicated, and therefore is by nature highly restricted to synchronous interactions.

The above remark means that it is possible to design an asynchronous messaging protocol, such as SMTP, supported by the Deep Random Secrecy primitives. None of the routing devices involved in the communication need to be sure, or even to know about Deep Random generation, they can just be storage and transmission devices.

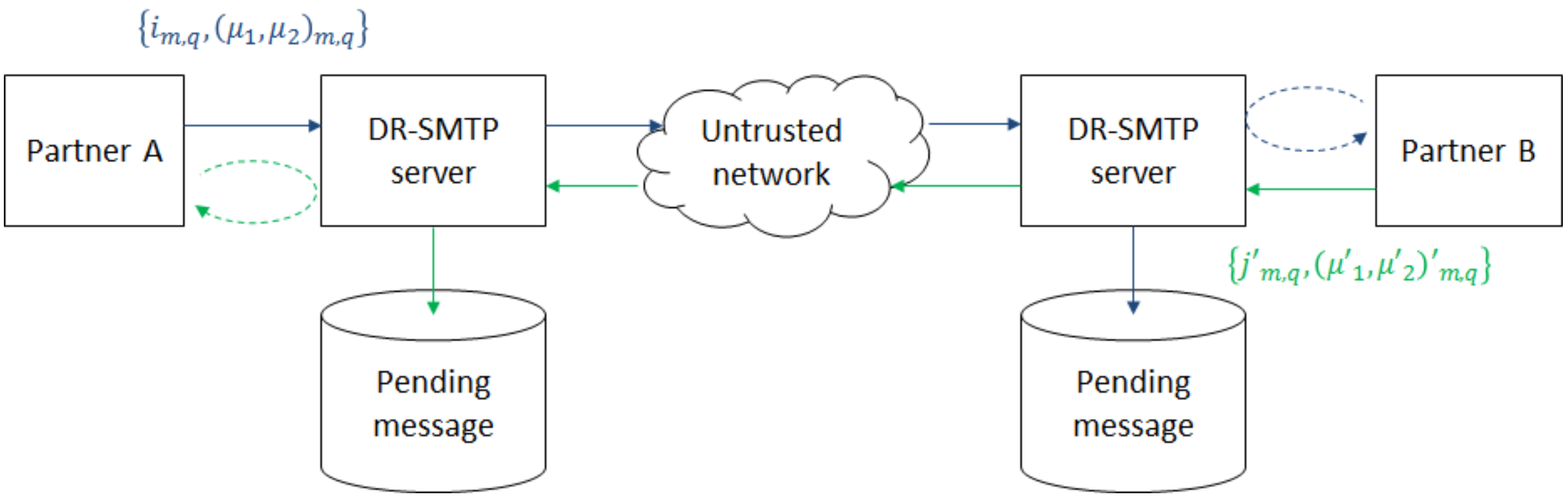

Fig. 5

## VI. Conclusion

We have proposed an extended set of protocols based on Deep Random Secrecy, benefiting from dramatic improvement in terms of bandwidth efficiency compared to previous presented versions. That improvement is based on reuse and recombination of bits, and we have justified that such reuse does not jeopardize the entropy of the distilled secret information.

We have provided new simulations, based on prior introduced methodology, showing in figures the increase of performance.

The bit reuse and recombine optimization technique has been applied to two primitives: (i) the unauthenticated protocol needing no prior shared knowledge between the partners and resisting to passive unlimited eavesdropper; and (ii) the authenticated protocol needing a short authentication shared secret between the partners, and resisting to active unlimited MITM.

We have also addressed consideration on practical design and implementation. We showed the different modules that should be comprised in a Deep Random Secrecy communication system, in particular, the DRG, and the Authentication Wallet. We have identified the important compromises between computation, memory and bandwidth that need to be considered, and we have discussed the various possibility of implementation for those modules, depending on the capacity of the host.

At last, we have also shown how one can combine the two primitives to implement more elaborated protocols that could be candidate to replace popular protocols such as TLS or SMTP in a post-quantum era.

We have discussed the fact that the performance of the overall protocol is also highly depending on the performance of the Reconciliation and Privacy Amplification techniques that we use. The primitives and simulations presented in this paper only use the basic ‘bit repetition’ reconciliation technique in order to show the emergence of Advantage Distillation phenomenon. Further work, in the perspective of practical implementation, should study how advanced Reconciliation techniques such as Cascade or LDPC codes can efficiently be introduced in the protocol to improve its efficiency.

**Who is the author ?**

I have been an engineer in computer science for 20 years. My professional activities in private sector are related to IT Security and digital trust, but have no relation with my personal research activity in the domain of cryptology. If you are interested in the topics introduced in this article, please feel free to establish first contact at tdevalroger@gmail.com

## Appendix I: Entropy of $H(x)_{\Phi_0}$

We have $H(x)_{\Phi_0} = \frac{n}{2} + O(\log n)$

■

$$H(x)_{\Phi_0} = 2\sum_{t=0}^{n/2}\sum_{x||x|=t}\frac{1}{\left(\frac{n}{2}+1\right)\binom{n/2}{t}}\log\left(\left(\frac{n}{2}+1\right)\binom{n/2}{t}\right) = \frac{2}{\left(\frac{n}{2}+1\right)}\sum_{t=0}^{n/2}\log\left(\left(\frac{n}{2}+1\right)\binom{n/2}{t}\right)$$

$$= \frac{2}{\left(\frac{n}{2}+1\right)}\sum_{t=0}^{n/2}\log\left(\binom{n/2}{t}\right) + O(\log n)$$

We have:

$$\sum_{t=0}^{n/2}\log\left(\binom{n/2}{t}\right) = \log\frac{\left((n/2)!\right)^{\frac{n}{2}+1}}{\left(\prod_{t=0}^{n/2} t!\right)^2} = \left(\frac{n}{2}+1\right)\log\left((n/2)!\right) - 2\log\left(\prod_{t=0}^{n/2} t!\right)$$

$$= \left(\frac{n}{2}+1\right)\log\left((n/2)!\right) - 2\sum_{u=1}^{n/2}\left(\frac{n}{2}+1-u\right)\log u$$

Injecting the two equalities below in the above,

$$\sum_{t=0}^{n/2}\log t = \frac{n}{2}\log\frac{n}{2} - \frac{n}{2} + O(1)$$

$$\sum_{t=0}^{n/2} t\log t = \frac{n^2}{8}\log\frac{n}{2} - \frac{n^2}{16} + O(n)$$

we eventually get the result. ■

## Appendix II: entropy in linear systems

For $p \in \mathbb{N}^*$, we consider the hyper-parallelepiped $S[0,p]$:

$$S[0,p]_{\{\lambda(r)\in\{0,1\}^n,\}}(x_1,\dots,x_n)\colon \left\{ x \in \mathbb{R}^n \,\middle|\, 0 \leq \sum_{s=1}^{n} \lambda_s(r)x_s \leq p \right\}_{1\leq r\leq n}$$

where we suppose that $\{\lambda(r) \in \{0,1\}^n\}$ is an independent linear system. For $\delta \in \mathbb{N}^*$, $\delta|p$, we also consider the hyper-cubic tiling of $\mathbb{R}^n$ defined by

$$\pi_{t_1,\dots,t_n}(x) = \left\{ x \in \mathbb{R}^n \,\middle|\, \forall r, \left\lfloor \frac{(\lambda(r)\cdot x - n\rho_r)}{\delta} \right\rfloor = t_r \right\}$$

where $\{\rho_r \in [0,1]\}$ are translation parameters. We eventually consider the checkerboard coloring function associated to that tiling:

$$e_r(x) = \left\lfloor \frac{(\lambda(r)\cdot x - n\rho_r)}{\delta} \right\rfloor \bmod 2$$

Our purpose in this section is to determine the entropy $H(\{e_1(x),\dots,e_n(x)\})$ depending on some statistical properties of the linear system $\{\lambda(r) \in \{0,1\}^n\}$ and the translation parameters $\{\rho_r \in [0,1]\}$.

**Proposition AII-1:**

*We consider the case where $|\lambda(r)| = p$, and $\delta|p$, and $\rho_r = 0\ \forall r$. If $x$ is equidistributed over $S[0,p]$, and if $p/\delta \equiv 0 \bmod 2$, then we have $H(\{e_1(x),\dots,e_n(x)\}) = n$*

■ As $p/\delta \equiv 0 \bmod 2$ and $p > 0$, there exists a vertex $(t_1,\dots,t_n)$ of the tiling that is contained in the interior strict of $S$. The linear system being independent, there are $2^n$ cells that are neighboring the vertex and each of them represents a unique combination of $\{e_1(x),\dots,e_n(x)\}$ in $\{0,1\}^n$.

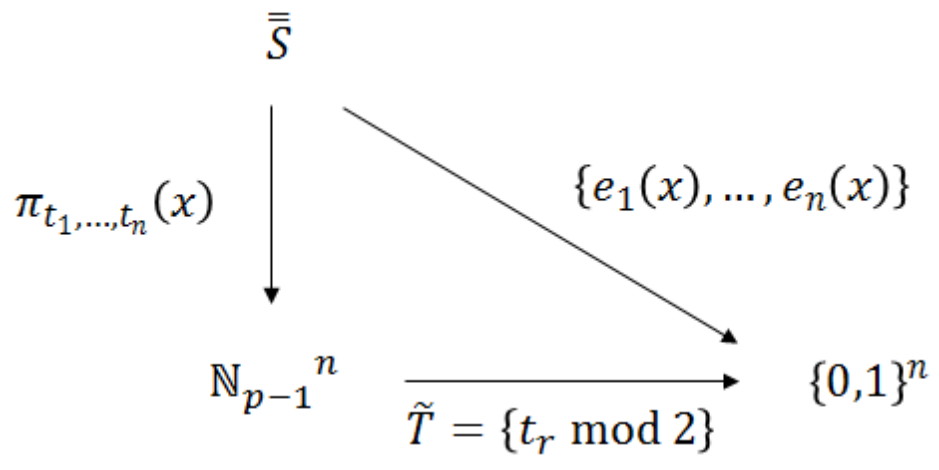

With the above diagram, one can see that this argument ensures, by $2\delta$ −translations in any dimension, that any combination of $\{e_1(x),\dots,e_n(x)\}$ in $\{0,1\}^n$ has the exact same number of inverse images by the equi-partitioning surjective function $\tilde{T}$. The distribution of $x$ being equidistributed over $S$, the above implies that $H(\{e_1(x),\dots,e_n(x)\}) = n$. ■

The above situation is fairly simple due to the fact that $x$ is equidistributed over $S[0,p]$. We consider now the more complex situation where $x$ is distributed over $[0,1]^n$, and $\{\rho_r\}$ are independent random variable, all being equidistributed over $[0,1]$. In order to simplify the computations, we will also make the reasonable assumption $(P)$ that, $\forall r$:

$$\frac{1}{n}\sum_{s=1}^{n} \int_{\{0,1\}^n} \lambda_s(r)\Phi_r(\lambda(r))d\lambda(r) = \theta \qquad (P)$$

**Proposition AII-2:**

*We consider the case where the system $\{\lambda(r)\}$ is a random linear system of independent vectors in $\{0,1\}^n$ also verifying $(P)$, and being statistically anisotropic; If $x$ is distributed over $[0,1]^n$, then we have*

$$H(\{e_1(x),\dots,e_n(x)\}) \approx n \min\left(1, \delta^n \sqrt{n!\,\theta^n(1-\theta)^{n-1}(n\theta-\theta+1)}\right)$$

■ By considering the statistical translations of the tiling $\pi_{t_1,\dots,t_n}(x)$, we obtain that the average number of solution $x \in [0,1]^n$ for the linear system: $\{e_1(x),\dots,e_n(x)\} = u \in \{0,1\}^n$ is given by

$$E\left[\frac{\mathrm{Vol}([0,1]^n)}{\mathrm{Vol}(\pi_{t_1,\dots,t_n})}\right]$$

Then obviously, $\mathrm{Vol}([0,1]^n) = 1$. By denoting $L$ the matrix associated to the linear system $\{\lambda(r)\}$, we can express the volume of the hyper-parallelepiped $\pi_{t_1,\dots,t_n}$ as:

$$\mathrm{Vol}(\pi_{t_1,\dots,t_n}) = \mathrm{Vol}(\pi_{0,\dots,0}) = \mathrm{Vol}(S[0,\delta]) = \int_{[0,1]^n} \left|Det\left(\frac{1}{\delta}L^{-1}\right)\right| dy = \frac{1}{\delta^n |DetL|}$$

By applying classical convexity inequality, we get:

$$E\left[\frac{\mathrm{Vol}([0,1]^n)}{\mathrm{Vol}(\pi_{t_1,\dots,t_n})}\right] \geq \frac{1}{E\left[\mathrm{Vol}(\pi_{t_1,\dots,t_n})\right]} \geq \delta^n (E[(DetL)^2])^{\frac{1}{2}}$$

We have of course

$$DetL = \sum_{\sigma \in \mathfrak{S}_n} \varepsilon(\sigma) \prod_{s=1}^{n} \lambda_s(\sigma(s))$$

and due to the fact that the vectors are independent and statistically anisotropic (which means that $\lambda(r)$ has the same probability than $\sigma(\lambda(r))$), it is equivalent, when computing $E[(DetL)^2]$, to assume that the matrix $L$ has independent binary random elements with parameter $\theta$. One can thus apply the result of lemma AII below, and get:

$$E\left[\frac{\mathrm{Vol}([0,1]^n)}{\mathrm{Vol}(\pi_{t_1,\dots,t_n})}\right] \geq \delta^n \sqrt{n!\,\theta^n(1-\theta)^{n-1}(n\theta-\theta+1)}$$

Eventually, the average number of solution $x \in [0,1]^n$ for the linear system: $\{e_1(x),\dots,e_n(x)\} = u \in \{0,1\}^n$ is lower bounded by

$$\delta^n \sqrt{n!\,\theta^n(1-\theta)^{n-1}(n\theta-\theta+1)}$$

■

The following lemma gives a proof for a variation of a well-known result:

**Lemma AII**

*Let L be a random matrix whose elements are independent binary random variables with parameter $\theta$. Then we have:*

$$E[(DetL)^2] = n!\,\theta^n(1-\theta)^{n-1}(n\theta-\theta+1)$$

■ We use the same combinatorial technique than in [9] Lemma 2. We denote, for $\sigma \in \mathfrak{S}_n$, $supp(\sigma) = \ker(\sigma - id_{\mathfrak{S}_n}) = \{i, \sigma(i) \neq i\}$ and $|\sigma| = card(supp(\sigma))$.

$$(DetL)^2 = \sum_{\sigma,\mu\in\mathfrak{S}_n} \varepsilon(\sigma)\varepsilon(\mu)\prod_{s=1}^{n}\lambda_s(\sigma(s))\lambda_s(\mu(s))$$

$$E[(DetL)^2] = \sum_{\sigma,\mu\in\mathfrak{S}_n} \varepsilon(\sigma)\varepsilon(\mu)\theta^{2(n-\#\{s\in\mathbb{N}_n^*|\sigma(s)=\mu(s)\})}\theta^{\#\{s\in\mathbb{N}_n^*|\sigma(s)=\mu(s)\}}$$

Reordering by using $\rho = \sigma^{-1}\circ\mu$, and seeing that $\varepsilon(\sigma)\varepsilon(\mu) = \varepsilon(\sigma\circ\mu) = \varepsilon(\sigma^2\circ\rho) = \varepsilon(\rho)$, we get:

$$E[(DetL)^2] = n!\,p^n \sum_{\rho\in\mathfrak{S}_n}\varepsilon(\rho)\theta^{|\rho|}$$

For $I$ a subset of $\mathbb{N}_n$ we define by $|I|$ the cardinality of $I$ and $\mathcal{S}_I$ the subgroup of $\mathfrak{S}_n$ including permutations for which $supp(\sigma) \subset I$. We will use the general inversion formula, for any function $f: \mathfrak{S}_n \mapsto \mathbb{R}$ :

$$F(s) \triangleq \sum_{\substack{\rho\in\mathfrak{S}_n\\|\rho|=s}} f(\rho)$$

$$G(s) \triangleq \sum_{\substack{I\\|I|=s}}\sum_{\rho\in\mathcal{S}_I} f(\rho)$$

We first establish a relationship between $G$ and $F$:

$$G(s) = \sum_{\substack{\rho\in\mathfrak{S}_n\\|\rho|\le s}}\left(\sum_{\substack{I\\|I|=s\\I\supset supp(\rho)}} 1\right)f(\rho) = \sum_{\substack{\rho\in\mathfrak{S}_n\\|\rho|\le s}}\binom{n-|\rho|}{s-|\rho|}f(\rho) = \sum_{u=0}^{s}\binom{n-u}{s-u}F(u)$$

This formula can be reversed by using the classical matrix inverse:

$$\begin{pmatrix} & \cdots & 0\\ \vdots & \ddots & \vdots\\ \binom{i}{j} & \cdots & \end{pmatrix}_{i,j\in\mathbb{N}_n}^{-1} = \begin{pmatrix} & \cdots & 0\\ \vdots & \ddots & \vdots\\ (-1)^{i+j}\binom{i}{j} & \cdots & \end{pmatrix}_{i,j\in\mathbb{N}_n}$$

(which can be easily seen by considering for instance the inverse of the $\mathbb{R}_n[X]$ endomorphism $P(X) \mapsto P(X+1)$).

By setting $j' = n - j$ and $i' = n - i$, we have $\binom{n-j}{i-j} = \binom{j'}{j'-i'} = \binom{j'}{i'}$, and thus:

$$\begin{pmatrix} & \cdots & 0 \\ \vdots & \ddots & \vdots \\ \binom{n-j}{i-j} & \cdots & \end{pmatrix}_{i,j\in\mathbb{N}_n}^{-1} = \begin{pmatrix} & \cdots & 0 \\ \vdots & \ddots & \vdots \\ (-1)^{i+j}\binom{n-j}{i-j} & \cdots & \end{pmatrix}_{i,j\in\mathbb{N}_n}$$

from which we take the reverse formula:

$$F(s) = \sum_{u=0}^{s} (-1)^{u+s} \binom{n-u}{s-u} G(u)$$

Thanks to the above, by choosing $f(\rho) = \varepsilon(\rho)$, we can write $E[(DetL)^2]$ depending on $G(u)$ as follows:

$$E[(DetL)^2] = n!\,\theta^n \sum_{s=0}^{n} \theta^s \sum_{u=0}^{s} (-1)^{u+s} \binom{n-u}{s-u} G(u) \qquad (lAII.1)$$

As soon as $m \geq 2$, $\mathfrak{S}_m$ contains a transposition $\tau$ and we have $\mathfrak{S}_m = \tau \circ \mathfrak{S}_m$, which implies that

$$\sum_{\rho\in\mathfrak{S}_m} \varepsilon(\rho) = 0$$

and therefore,

$$G(u) = \sum_{\substack{I \\ |I|=u}} \sum_{\rho\in\mathcal{S}_I} \varepsilon(\rho) = \begin{cases} 1 \; if \; u = 0 \\ n \; if \; u = 1 \\ 0 \; if \; u \geq 2 \end{cases}$$

Injecting the above in $(lAII.1)$ we get:

$$E[(DetL)^2] = n!\,\theta^n \sum_{s=0}^{n} \theta^s \left( (-1)^s \binom{n}{s} + (-1)^{s+1} n \binom{n-1}{s-1} \right) = n!\,\theta^n \sum_{s=0}^{n} \theta^s (-1)^{s+1} (s-1) \binom{n}{s}$$
$$= n!\,\theta^n \left( (1-\theta)^n + n\theta(1-\theta)^{n-1} \right) = n!\,\theta^n (1-\theta)^{n-1} (n\theta - \theta + 1)$$

■

## Appendix III: Simulations for the optimized protocol

### Handling dummy DRGs

In our simulation, we don't focus on DRG. The DRG of each partner is emulated from the fixed distribution $\Phi_0$ in $\zeta(\alpha)$ with $\sigma_{\Phi_0} = \mathrm{Id}_{\mathfrak{S}_n}$.

### Input parameters, dispersion and error correcting

The simulation is executed for a given set of parameters $(n, k, K, L)$ and, at each execution, it repeats $R$ times the protocol. The dispersion is simulated by the generation of $\sigma_d[i]$ (resp. $\sigma'_d[j]$) in such a way that $i \subset \sigma_d[i](I_0)$ (resp. $j \subset \sigma'_d[j](I_0)$) and randomly, one time over two in average, $A$ (resp. also and independently for $B$) chooses $\sigma_A = \sigma'_d[j]$ rather than $\sigma_A = \mathrm{Id}$ (resp. $\sigma_B = \sigma_d[i]$ rather than $\sigma_B = \mathrm{Id}$), which simulates the random choice of $A$ (resp. also and independently for $B$) among $(\mu'_1, \mu'_2)$ (resp. $(\mu_1, \mu_2)$). One can notice that it is useless to introduce random permutation in the draw of $\Phi_0$ because it is easy to verify that $\sigma_d[i]_{[\Phi_0 \circ \mu]} = \mu^{-1} \circ \sigma_d[i]_{[\Phi_0]}$ and thus

$$\Phi_0 \circ \mu \circ \sigma_d[i]_{[\Phi_0 \circ \mu]} = \Phi_0 \circ \sigma_d[i]_{[\Phi_0]}$$

In those simulations, we mainly focus on Advantage Distillation. The effect of Privacy Amplification will be discussed for one of the simulations. Information Reconciliation is not fully considered, but basic error correcting methods are handled at step 6 in order to lower the error rate at a level where efficient methods (LDPC, Cascade) can operate. We will in particular cumulate, sequentially two error correcting methods based on bit repetition:

1) Bit repetition with majority decoding: the codewords $v_A$ chosen by $A$ can only be $(0,0,\ldots,0)_L$ or $(1,1,\ldots,1)_L$ depending on $e_A = 0$ or $e_A = 1$. $B$ decodes $e_B = 0$ if there is a majority of 0 in the received sequence, and decodes $e_B = 1$ otherwise. This code is suitable to lower high error rates (initially close to 50%) at a reasonable bandwidth cost.
2) Bit repetition with exact decoding: the codewords $v_A$ chosen by $A$ can only be $(0,0,\ldots,0)_L$ or $(1,1,\ldots,1)_L$ depending on $e_A = 0$ or $e_A = 1$. $B$ publicly discards all decoded sequence $v_B$ that is not $(0,0,\ldots,0)_L$ or $(1,1,\ldots,1)_L$ and obviously decodes accordingly $e_B = 0$ if $|v_B| = 0$, and $e_B = 1$ if $|v_B| = L$. This code has a much better acceleration effect on reducing the error rate as soon as the initial error rate is significantly lower than 50%; however, more sophisticated codes (LDPC, Cascade) would do an even better job.

### Behavior of the opponent

The behavior of the opponent is tested with 2 canonical strategies among $\Omega_\#(G, \mathcal{P})$ and $\Omega'_\#(G, \mathcal{P})$:

$$\omega_1(i, j) = \frac{k|i||j|}{n^2}$$

$$\omega_2(i, j, (\mu_1, \mu_2), (\mu'_1, \mu'_2)) = \frac{2k\left(\left(\sum_{r \in \sigma_{\xi,B}(I_0)} i_r\right)\left(\sum_{r \in \sigma_{\xi,A}(I_0)} j_r\right) + \left(\sum_{r \in \overline{\sigma_{\xi,B}(I_0)}} i_r\right)\left(\sum_{r \in \overline{\sigma_{\xi,A}(I_0)}} j_r\right)\right)}{n^2}$$

where $\sigma_{\xi,B}$ is a permutation randomly chosen by $\xi$ among $(\mu_1, \mu_2)$ and $\sigma_{\xi,A}$ is a permutation randomly chosen by $\xi$ among $(\mu'_1, \mu'_2)$. The relevance of the choice of $\omega_2$ is justified in [9] Proposition 9. However, the discarding action at Advantage Distillation step 5, makes strategy $\omega_2$ totally useless (meaning leading to an opponent's error rate of exactly $1/2$). Anyway, it is shown in [9] that any strategy is almost useless (meaning leading to an opponent's error rate of $1/2 - o(1)$). Strategy $\omega_1$ is adopted in the results of simulations presented below.

**Output of simulation**

Each execution of the simulation outputs 3 elements of information:

(i) the net error rate for the legitimate partners – denoted $\varepsilon$,
(ii) the net knowledge rate for the opponent – denoted $\varepsilon'$, and
(iii) a lower bound of Cryptologic Limit – denoted $CL$.

$P(e_A \neq e_B)$ and $\left|P(e_E = e_B) - \frac{1}{2}\right|$ can be seen respectively as the error rate and 'eavesdropping' rate, so both with values in $[0,1]$. We decided in [9], by convention that fits the common sense, that $P(e_A \neq e_B)$ cannot be greater than $1/2$, and $P(e_E = e_B)$ cannot be lower than $1/2$; indeed, if $P(e_A \neq e_B) > 1/2$ one can replace $e_B$ by $\overline{e_B}$ (we then denote ${e_B}^*$ the final choice for $e_B$), and if $P(e_E = e_B) < 1/2$ one can replace $e_E$ by $\overline{e_E}$ (we then denote ${e_E}^*$ the final choice for $e_E$). And therefore, we can define $\varepsilon$ as the net error rate, and $\varepsilon'$ as the net knowledge rate (by the opponent), with the following formulas:

(i) $\varepsilon = 2\left(\min\left(P(e_A \neq e_B), 1 - P(e_A \neq e_B)\right)\right) = 2P(e_A \neq {e_B}^*)$

(ii) $\varepsilon' = 2\left(\max\left(P(e_E = e_B), 1 - P(e_E = e_B)\right) - \frac{1}{2}\right) = 2\left(P(e_A = {e_E}^*) - \frac{1}{2}\right) > 0$

$CL$ as introduced in [9], represents the rate of perfectly reliable bit transmitted over the total amount of bit transmitted, which in the case of the protocol is:

(iii) $CL = \frac{L^*}{6n^2}(1 - \varepsilon - \varepsilon')$ (one can remark that it is possible to only transmit $\mu_x(I_0)$ instead of $\mu_x$, which can be coded with $n$ bits instead of $n \log_2 n$).

The 3 output information are measured as averaged over all the $R$ repetitions of the protocol within an execution of the simulation.

**Presentation of the results**

We have executed series of simulations with variations of parameters $(n, K, L)$, and with $R$ being sufficiently large so that the residual quantity of distilled digit is at least $L^* > 100$.

In the following results, the sampling parameter $K$ is expressed as a multiple of $\frac{1}{2\sqrt{nk}}$.

In all the graphics below, the abscissa represents the coursing parameter values, the left ordinate represents the values of error rate $\varepsilon$ and knowledge rate $\varepsilon'$, the right ordinate represents the values of the Cryptologic Limit's lower bound $CL$, and the title of the graphic gives the values of the fixed parameters, and the range of the coursing parameter.

The first graphic below presents Advantage Distillation without the improving effects of error correcting codes. We only increase the length of the Bernoulli vector $n$ from 5,000 to 50,000. The other fixed parameters are $k = 12$, $K = 12$, $w = 700$:

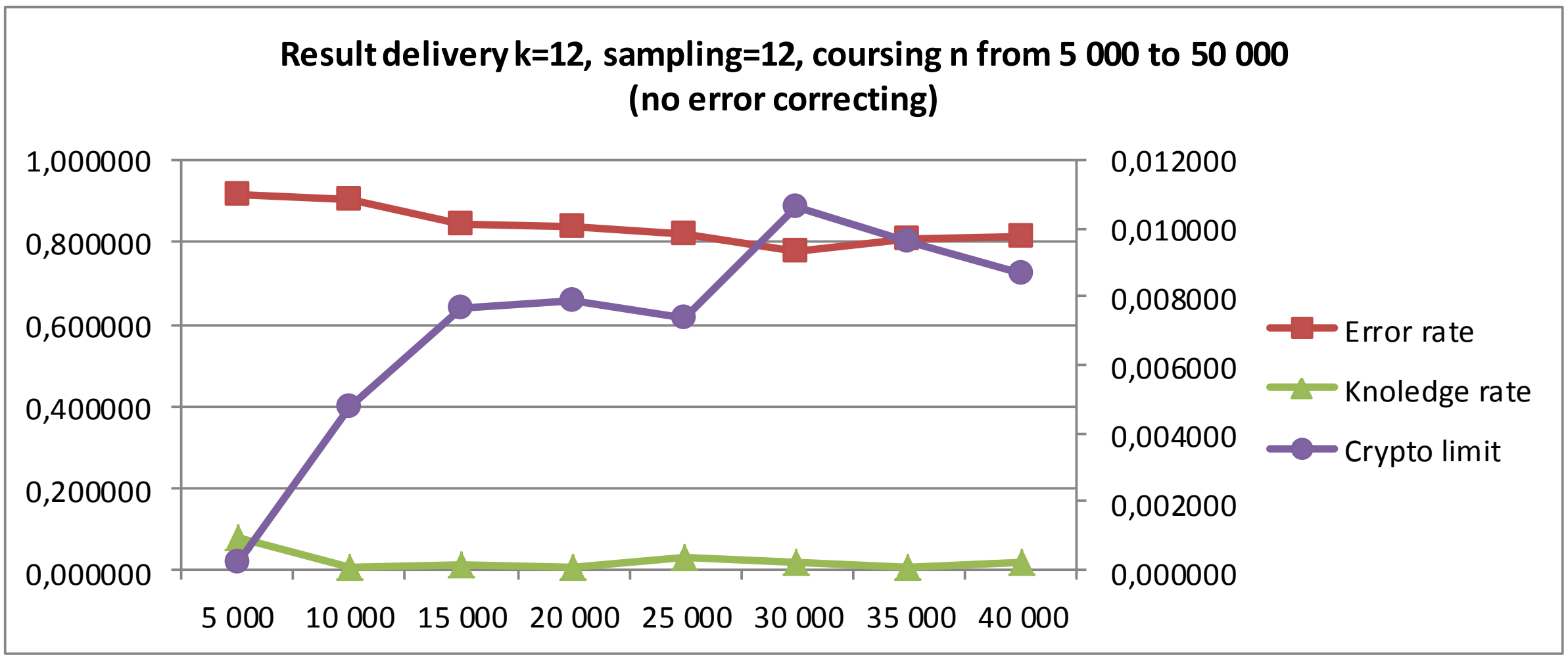

We observe that the error rate is decreasing when $n$ increases. This can be explained by the fact that it is needed to have large values of $n$ in order to have a fair occurrence probability that condition $|T_m| = 2$ and condition ($\sigma_B = \sigma_\Phi$ and $\sigma_A = \sigma_{\Phi'}$) above coexist. Nevertheless, the error rate of the partners remains very high (while, as expected, the knowledge rate is remains almost null), even if Advantage Distillation is obtained. In terms of bandwidth, we have adopted a reuse parameter $w = 700 < n$, which makes $CL$ non optimal, but was needed to remain in the limit of the available RAM for that simulation.

The second graphic below presents the impact of error correcting. The 'Bit repetition with majority decoding' uses a fix length of codewords equal to 31, while we increase the value of the 'Bit repetition with exact decoding' codewords' length from 1 to 10. The other fixed parameters are $= 30{,}000$ $k = 12$, $K = 12$, $w = 1{,}000$:

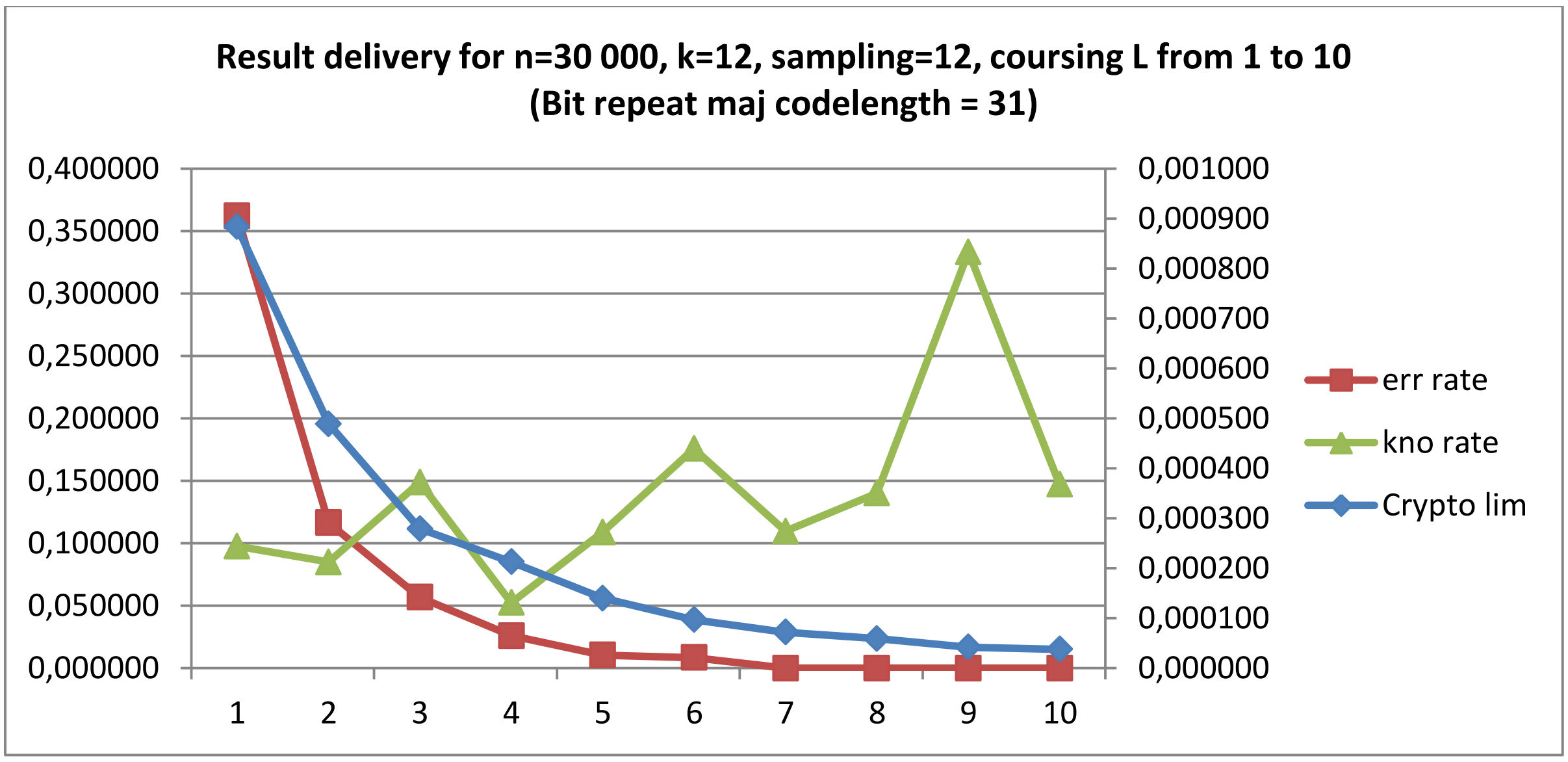

We observe that we manage to obtain an accelerated decrease of the error rate thanks to the (non-optimal) 'Bit repetition with exact decoding' technique. It is of course at cost of increasing the knowledge rate (which can be then rectified by Privacy Amplification). The Cryptographic limit decreases due to the increasing discarding attached to that technique.

The third graphic presents the effect of Privacy Amplification. We apply a simple Privacy Amplification method corresponding to the universal hashing function « multiply-add-shift » $h_{a,b}(x) = \left((ax+b) \bmod 2^{PA}\right) \operatorname{div} 2^{PA-1}$, for which parameters $a, b$ are picked for each new output digit (distilled from sequences of $PA$ input digits). The codewords size $PA$ varies from 2 to 12. The other fixed parameters are $n = 10{,}000$ $k = 12$, $K = 12$, $w = 3{,}000$, the 'Bit repetition with majority decoding' uses a fix length of codewords equal to 31, and the 'Bit repetition with exact decoding' codewords' uses a fix length of codewords equal to 6:

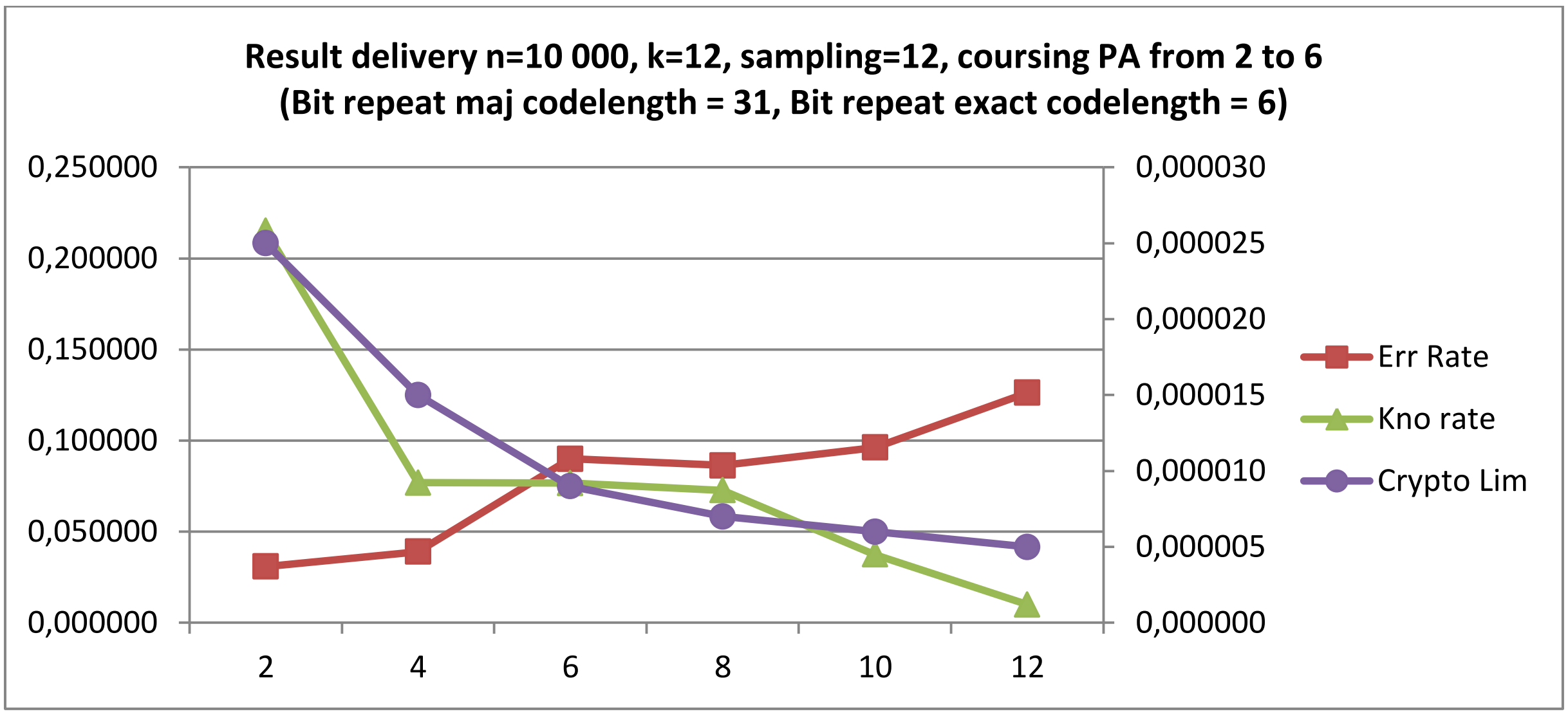

The reduction of the knowledge rate is of course at cost of the increase of the error rate, and at the overall decrease of $CL$. This simulation is not practically relevant, because Privacy Amplification should be used only after a more efficient Reconciliation method is used to lower the error rate; it is just purposed to show the qualitative effect of Privacy Amplification in the context of a simplistic version of the Deep Random secrecy protocol.